%% file: master.tex
\pdfoutput=1
\documentclass[11pt, draftclsnofoot, onecolumn]{IEEEtran}

\ifCLASSINFOpdf

\else

\fi

\usepackage{fancyhdr}
\usepackage{amsfonts}
\usepackage{amssymb}
\usepackage[utf8]{inputenc}
\usepackage[pdftex]{graphicx}
\DeclareGraphicsExtensions{.pdf,.jpeg,.png,.eps}
\usepackage[cmex10]{amsmath}
\usepackage{color}
\usepackage{algorithm}
\usepackage{algorithmic}
\usepackage{amsthm}
\usepackage{cases}
\newtheorem{theorem}{Theorem}
\newtheorem{lemma}{Lemma}

\newtheorem{definition}{Definition}

\usepackage{comment}
\usepackage{setspace}
\usepackage{subfigure}
\usepackage[font={small}]{caption}
\usepackage[numbers,sort&compress]{natbib}
\usepackage{epstopdf}

\makeatletter
\renewcommand\appendix{\par
  \setcounter{section}{0}
  \setcounter{subsection}{0}
  \setcounter{figure}{0}
  \setcounter{table}{0}
  \renewcommand\thesection{Appendix \Alph{section}}
  \renewcommand\thefigure{\Alph{section}\arabic{figure}}
  \renewcommand\thetable{\Alph{section}\arabic{table}}
}
\makeatother

\begin{document}

\pagestyle{fancy}
\rhead{\textit{IEEE Transactions on Wireless Communications}}

%
\title{Defeating jamming with the power of silence: \\ a game-theoretic analysis}
\author{Salvatore D'Oro,~\IEEEmembership{Student Member, IEEE,} 
	Laura Galluccio,~\IEEEmembership{Member, IEEE,}
        Giacomo Morabito,
        Sergio Palazzo,~\IEEEmembership{Senior Member, IEEE,}
	Lin Chen,~\IEEEmembership{Member, IEEE,}
	and Fabio Martignon,~\IEEEmembership{Member, IEEE,}\\ 
\thanks{\textbf{IEEE copyright notice:} This paper has been accepted for publication in the IEEE Transactions on Wireless Communications. 
“© © 2014 IEEE. Personal use of this material is permitted. Permission from IEEE must be obtained for all other uses, 
in any current or future media, including reprinting/republishing this material for advertising or promotional purposes, creating new collective works,
for resale or redistribution to servers or lists, or reuse of any copyrighted component of this work in other works.”}
\thanks{S. D'Oro, L. Galluccio, G. Morabito and S. Palazzo are with the CNIT Research Unit at University of Catania, Italy  (e-mail: \{name.surname\}@dieei.unict.it).
L. Chen is with the Laboratoire de Recherche en Informatique (LRI), Université Paris-Sud, Paris, France (e-mail: Lin.Chen@lri.fr).
F. Martignon is with the Laboratoire de Recherche en Informatique (LRI), Université Paris-Sud, Paris, France, and is also with the Institut Universitaire de France (IUF), (e-mail: fabio.martignon@lri.fr).}
\thanks{This work has been supported by the European Commission in the framework of the FP7 Network of
Excellence in Wireless COMmunications NEWCOM\# (Grant agreement no. 318306) and partially supported by the French ANR in the framework of the Green-Dyspan project.
}}

\maketitle

\vspace{-2cm}
\begin{abstract}
The timing channel is a logical communication channel in which information is encoded in the timing between events. 
Recently, the use of the timing channel has been proposed as a countermeasure to reactive jamming attacks performed by an energy-constrained malicious node.
In fact, whilst a jammer is able to disrupt the information contained in the attacked packets, timing information cannot be jammed and, therefore, timing channels can be
exploited to deliver information to the receiver even on a jammed channel.

Since the nodes under attack and the jammer have conflicting interests, their interactions can be modeled by means of game theory.
Accordingly, in this paper a game-theoretic model of the interactions between nodes exploiting the timing channel to achieve resilience to
jamming attacks and a jammer is derived and analyzed.
More specifically, the Nash equilibrium is studied in the terms of existence, uniqueness, and convergence under best response dynamics.
Furthermore, the case in which the communication nodes set their strategy and the jammer reacts accordingly is modeled and analyzed as a Stackelberg game, 
by considering both perfect and imperfect knowledge of the jammer's utility function.
Extensive numerical results are presented, showing the impact of network parameters on the system performance.
\end{abstract}


\input{Introduction}
\input{RelatedWork}
\input{GameGiacomo}
\input{NE_SCMGiacomo3_short_appendix}
\input{Numerical}
\input{ConcBib}

\input{bib.bbl}

\footnotesize
\bibliographystyle{IEEEtran}
\bibliography{IEEEabrv}

\section{Appendices}
\appendix
\input{appendixit}

\vspace{-1cm}
\begin{IEEEbiography}
    [{\includegraphics[width=1in,height=1.25in,keepaspectratio]{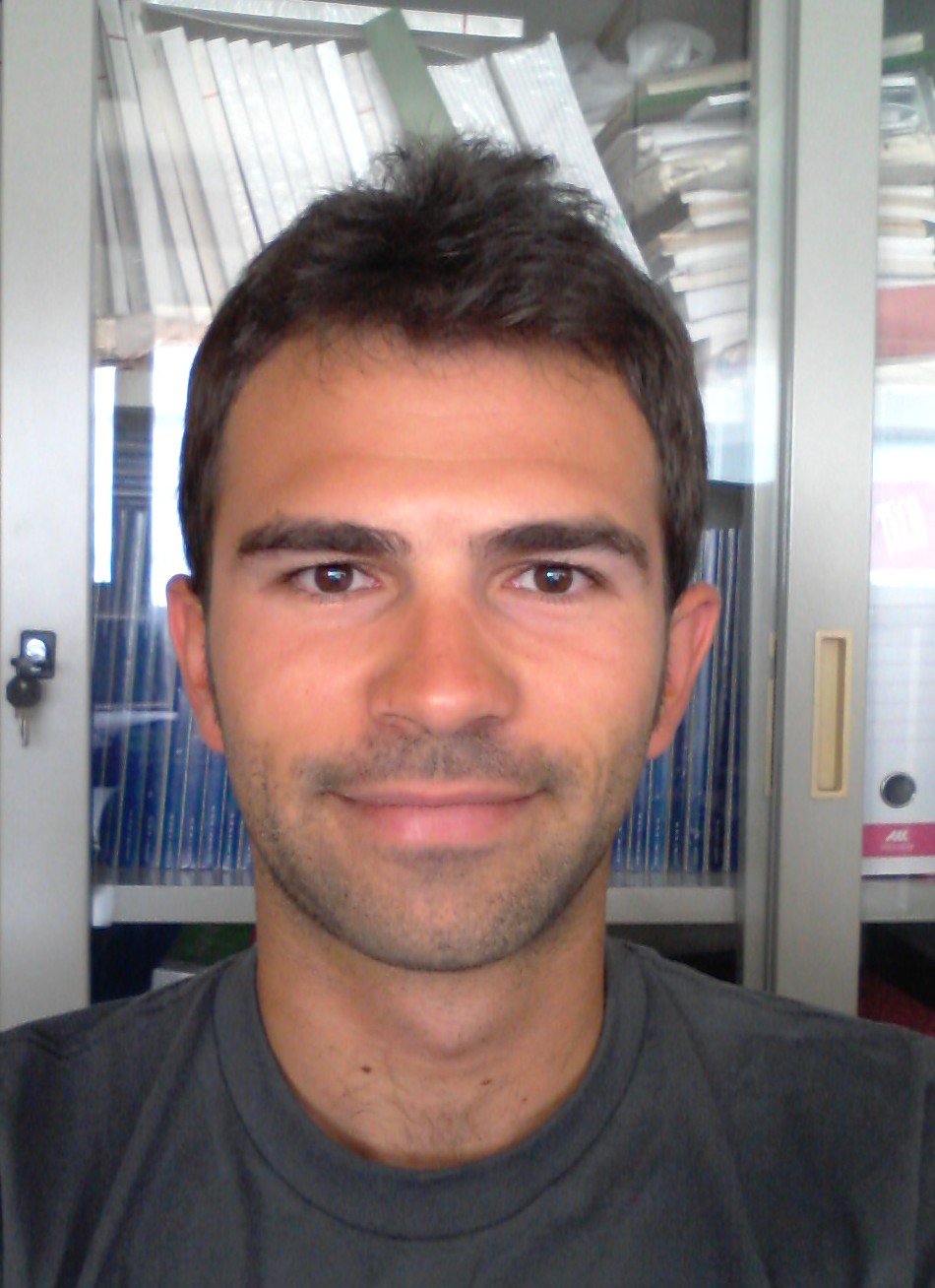}}]{Salvatore D'Oro} (S'12)
received the B.S. degree in Computer Engineering and the M.S. degree in Telecommunications Engineering degree both at the University of Catania.
In 2013, he spent a period as a Visiting Researcher at Centre National de la Recherche Scientifique (CNRS), Université Paris-Sud 11, Paris, France.
 He is currently a PhD Student at the University of Catania and his main interests are next generation communication networks, security and game theory.
In 2013 served on the Technical Program Committee (TPC) of the 20th European Wireless Conference (EW2014).
\end{IEEEbiography}

\vspace{-1cm}

\begin{IEEEbiography}
		[{\includegraphics[width=1in,height=1.25in,keepaspectratio]{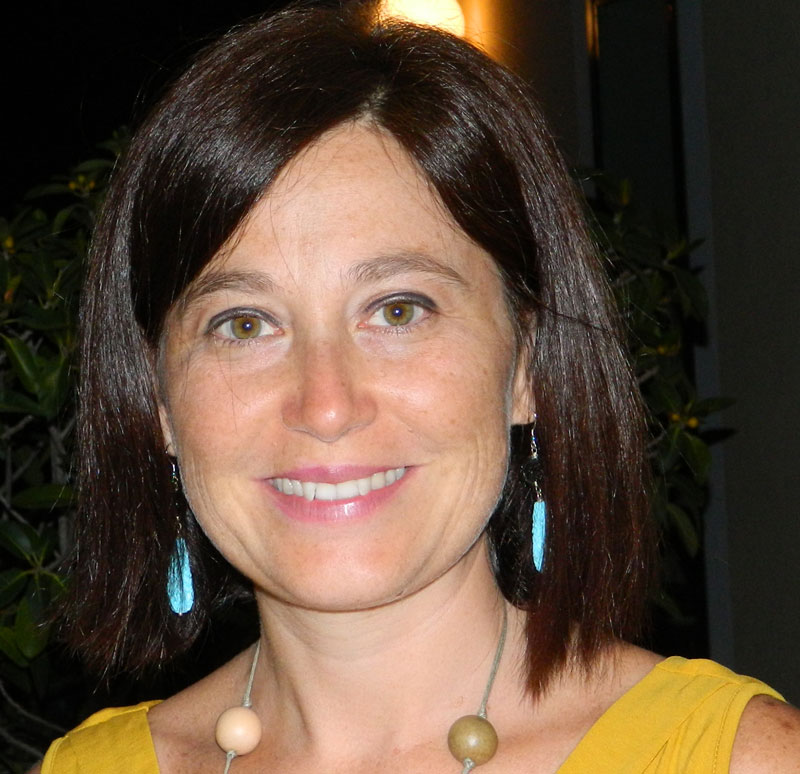}}]{Laura Galluccio}  (M'02) received the Laurea Degree in electrical engineering in 2001 and the Ph.D. degree in electrical, computer, and telecommunications engineering in 2005 from the University of Catania,  Italy. From 2002 to 2009 she was with the Italian National Consortium of Telecommunications (CNIT), working as a Research Fellow in the FIRB VICOM and NoE SATNEX projects.  Since 2010, she has been an Assistant Professor with the University of Catania. In 2005 she
was Visiting Scholar with the COMET Group, Columbia University, New York. Her research interests include wireless networking, unconventional communication networks, and network performance analysis. She serves in the editorial boards of Elsevier Ad Hoc Networks and Wiley Wireless Communications and Mobile Computing journals.
\end{IEEEbiography}

\vspace{-1cm}

\begin{IEEEbiography}
		[{\includegraphics[width=1in,height=1.25in,keepaspectratio]{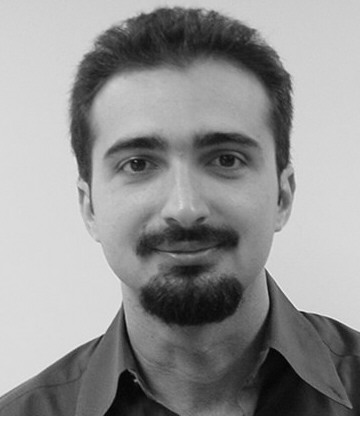}}]{Giacomo Morabito} received the laurea degree and the PhD in Electrical, Computer and Telecommunications Engineering from the Istituto di Informatica e Telecomunicazioni, University of Catania in 1996 and 2000, respectively. From November 1999 to April 2001, he was with the Broadband and Wireless Networking Laboratory of the Georgia Institute of Technology as a Research Engineer. Since April 2001 he is with the Dipartimento di Ingegneria Informatica e delle Telecomunicazioni of the University of Catania where he is currently Associate Professor. His research interests focus on analysis and solutions for wireless networks and Internet of Things.
\end{IEEEbiography}

\vspace{-1cm}

\begin{IEEEbiography}
		[{\includegraphics[width=1in,height=1.25in,keepaspectratio]{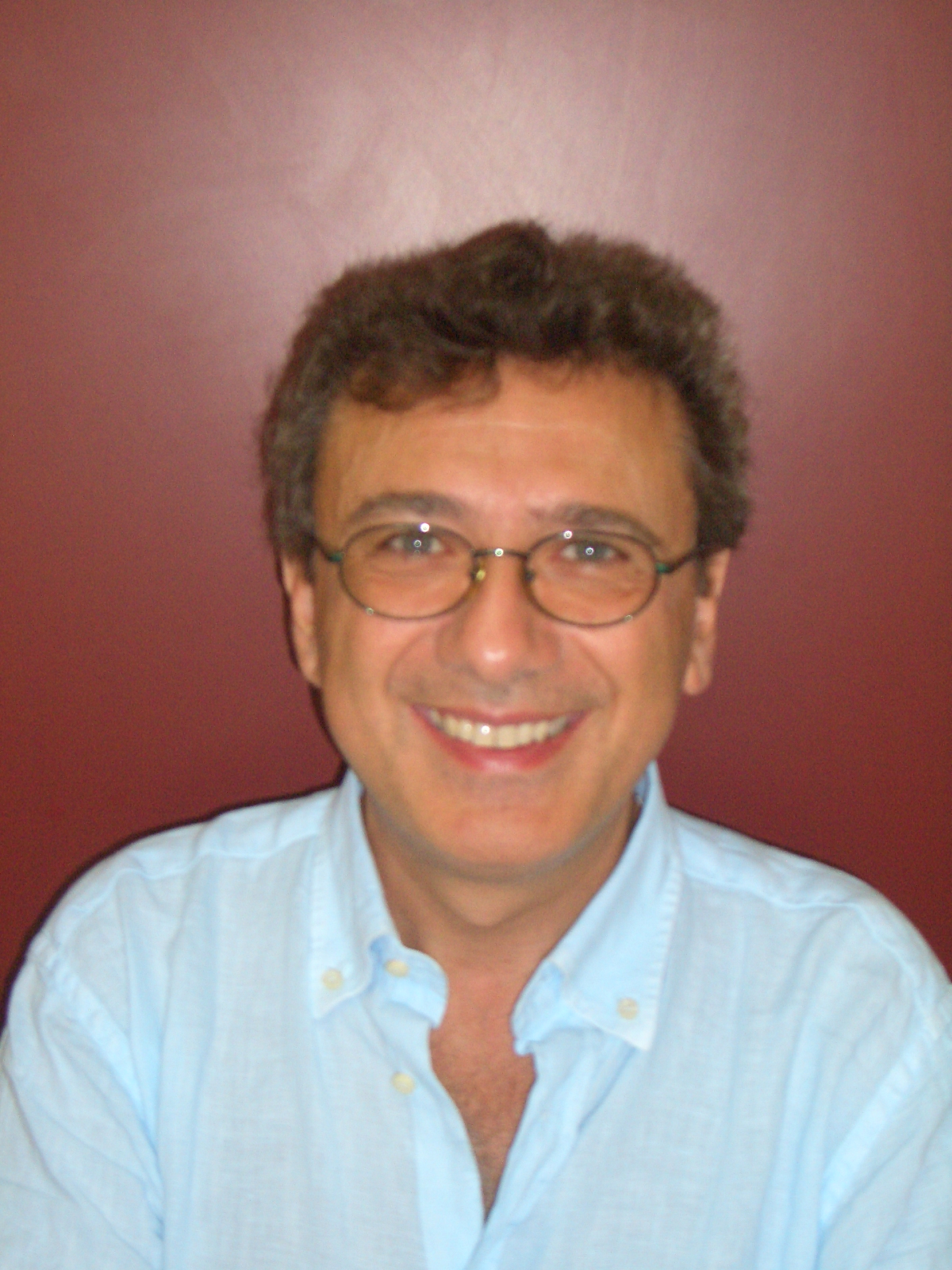}}]{Sergio Palazzo} (M’92$-$SM’99) received the degree
in electrical engineering from the University of
Catania, Catania, Italy, in 1977.
Since 1987, he has been with the University of
Catania, where is now a Professor of telecommunications
networks. His current
research interests include mobile systems, wireless
and satellite IP networks, and protocols for the next
generation of the Internet.
He has been the General Chair of some ACM conferences, including MobiHoc
2006 and MobiOpp 2010, and currently is a member of the MobiHoc Steering
Committee. He has also been the TPC Co-Chair of some other conferences, including
IFIP Networking 2011, IWCMC 2013, and EuropeanWireless 2014. He
currently serves on the Editorial Board of Ad Hoc Networks. In the recent past,
he also was an Editor of the IEEE Wireless Communications Magazine, IEEE/ACM Transactions
on Netrowking, IEEE Transactions on Mobile Computing, Computer
Networks, and Wireless Communications and Mobile Computing. 
\end{IEEEbiography}

\vspace{-1cm}

\begin{IEEEbiography}
		[{\includegraphics[width=1in,height=1.25in,keepaspectratio]{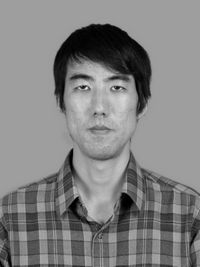}}]{Lin Chen} received his B.E. degree in Radio Engineering from
Southeast University, China in 2002 and the Engineer Diploma from 
Telecom ParisTech, Paris in 2005. He also holds a M.S. degree of 
Networking from the University of Paris 6. 
He currently works as associate professor in the department of 
computer science of the University of Paris-Sud. His main research interests include modeling and control for wireless 
networks, security and cooperation enforcement in wireless networks and 
game theory.
\end{IEEEbiography}

\vspace{-1cm}

\begin{IEEEbiography}
[{\includegraphics[width=1in,height=1.25in,clip,keepaspectratio]{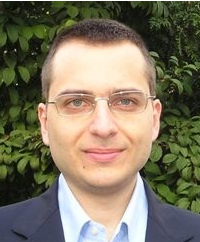}}] {Fabio Martignon} received the M.S. and the Ph.D. degrees in telecommunication engineering from the Politecnico di Milano in October 2001 and May 2005, respectively. He has been associate professor at University of Bergamo, and he is now full professor
in LRI (Laboratory for Computer Science) at Paris Sud University, and member of Institut Universitaire de France. His current research activities include cognitive radio networks, content-centric networks, network planning and game theory applications to networking problems.
\end{IEEEbiography} 

\vspace{-1cm}

\end{document}

%% file: Introduction.tex
\section{Introduction}

A \emph{timing channel} is a communication channel which exploits silence intervals between consecutive transmissions to encode information \cite{Anantha96}.
Recently, use of timing channels has been proposed in the wireless domain to support low rate, energy efficient communications \cite{Morabito11,Morabito13} 
as well as covert and resilient communications \cite{wisec, DOro2013}.

In this paper we focus on the resilience of timing channels to jamming attacks \cite{jampaper,survey}.
In general, these attacks can completely disrupt communications when the jammer continuously emits a high power disturbing signal, i.e., when \emph{continuous jamming} is performed.
However, continuous jamming is very costly in terms of energy consumption for the jammer \cite{book,eneff1,eneff2}.
This is the reason why in most scenarios characterized by energy constraints for the jammer, e.g., \textcolor{black}{when} the jammer is battery powered, 
non continuous jamming such as \emph{reactive jamming} is considered.
In this case the jammer continuously listens over the wireless channel and begins the transmission of a high power disturbing signal as soon
as it detects \textcolor{black}{an ongoing} transmission activity.
Effectiveness of reactive jamming has been demonstrated and its energy cost analyzed in \cite{Wilhelm, eneff2, jampaper, Xu}.

Timing channels are more - although not totally \cite{wisec} - immune from reactive jamming attacks.
In fact, the interfering signal begins its disturbing action against the communication only after \textcolor{black}{identifying an ongoing transmission, and thus after} the timing information has been decoded 
by the receiver.
In \cite{wisec}, for example, a timing channel-based communication scheme has been proposed \textcolor{black}{to counteract jamming by establishing}  a low-rate physical layer on 
top of the traditional physical/link layers using detection and timing of failed packet receptions at the receiver.
In \cite{DOro2013}, instead, the energy cost of jamming the timing channel and the resulting trade-offs have been analyzed.

In this paper we analyze the interactions between the jammer and the node whose transmissions are under attack, which we call \emph{target node}.
Specifically, we assume that the target node wants to maximize the amount of information that can be transmitted per unit of time by means 
of the timing channel\footnote{Note that in this context energy is  not a concern for the target 
node, since by exploiting the timing channel, a significant reduction in the energy consumption can be obtained as demonstrated in \cite{Morabito11}.
}, whereas, the jammer wants to minimize such amount of information while reducing the energy expenditure\footnote{\textcolor{black}{Up to now, despite the wide literature in this context, a universal model describing how jammers and target nodes behave in real adversarial scenarios is missing. Therefore, in our study we tried to propose a high-level model that describes rational and realistic behavior of each player, by considering several elements that are related to hardware parameters and the energy/power concerns.}}.
As the target node and the jammer have conflicting interests, we develop a game theoretical framework that models their interactions.
We investigate both the case in which these two adversaries play their strategies simultaneously, and the \textcolor{black}{situation when}
the target node (the leader) anticipates the actions of the jammer (the follower).
To this purpose, we study both the Nash Equilibria (NEs) and Stackelberg Equilibria (SEs) of our proposed games. 

The main contributions of this paper can be therefore summarized as follows:
1) we model the interactions between a jammer and a target node as a jamming game;
2) we \textcolor{black}{prove} the existence, uniqueness and convergence to the Nash equilibrium (NE) under best response dynamics;
3) we prove the existence and uniqueness of the equilibrium of the Stackelberg game where the target node plays 
as a leader and the jammer reacts consequently;
4) we investigate in this latter Stackelberg scenario the impact on the achievable performance of \textit{imperfect knowledge} of the jammer's utility function;
5) we conduct an extensive numerical analysis which shows that our proposed models well capture the main factors behind the utilisation of timing channels, 
thus representing a promising framework for the design and understanding of such systems.

Accordingly, the rest of this paper is organized as follows. 
Related work is presented in Section \ref{RelatedWork}.
\textcolor{black}{In Section \ref{game} the proposed jamming game model is presented}. 
A theoretical study of the existence and uniqueness of the NE as well as of the convergence of the game to that equilibrium under best response dynamics is derived 
in Section \ref{NE_SCM}. Existence and uniqueness of the SE are discussed in Section \ref{SE_SCM}, together with some considerations relevant to 
imperfect knowledge scenarios. Then, numerical results are illustrated in Section \ref{numericalres}. 
Finally, in Section \ref{conclusions} conclusions are drawn.

%% file: RelatedWork.tex
\section{Related Work}  \label{RelatedWork}

Wireless networks are especially prone to several attacks due to the shared and broadcast nature of the wireless medium.
One of the most critical attacks is \emph{jamming} \cite{jampaper,survey}.
Jamming attacks can partially or totally disrupt ongoing communications, and proper solutions have been proposed in various application scenarios \cite{jampaper, eneff1, eneff2}.
\textit{Continuous} jamming attacks can be really expensive \textcolor{black}{for the jammer} in terms of energy 
consumption as the transmission of jamming signals needs a significant, and constant, amount of power.
To reduce energy consumption while achieving a high jamming effectiveness, \textit{reactive} jamming is frequently used \cite{Xu, Strasser10, Wilhelm, DOro2013}.
In \cite{Xu} and \cite{Strasser10} the feasibility and detectability of jamming attacks in wireless networks are analyzed. 
In \textcolor{black}{these papers above}, methodologies
to detect jamming attacks are illustrated; it is also shown that it is possible to identify which kind of jamming attack is ongoing 
by looking at the signal strength and other relevant network parameters, such as bit and packet errors.
In \cite{Wilhelm} Wilhelm et al. investigate  the feasibility of reactive jamming attacks by providing a real implementation of a reactive jammer in a
software-defined radio environment where a reactive jammer prototype is implemented on a USRP2 platform and network users are implemented on MICAz motes.
Authors show that reactive jamming attacks are feasible and efficient, and that low reaction times can be achieved;
then, they highlight the need to investigate proper countermeasures against reactive jamming attacks.

Several solutions against reactive jamming have been proposed that exploit different techniques,
such as frequency hopping \cite{Strasser08, Wang}, power control \cite{Yang} and unjammed bits \cite{Liu12} (see \cite{survey, jampaper} for surveys).
However, such solutions usually \textcolor{black}{rely on} users' cooperation and coordination, which might not be guaranteed in a jammed environment.
\textcolor{black}{In fact, the reactive jammer can } totally
disrupt each transmitted packet and, \textcolor{black}{consequently}, no information can be decoded and then used to this purpose.

\emph{Timing channels} have been frequently exploited to support covert low rate \cite{Anantha96},
energy efficient \cite{Morabito11,Morabito13} and undetectable communications \cite{Liu}. Also, they have been proposed as anti-jamming solutions \cite{wisec, DOro2013}.
More specifically, in \cite{wisec} Xu et al. propose an anti-jamming timing channel that exploits 
inter-arrival times between jammed packets to encode information to be transmitted, showing how timing channels are 
suitable to guarantee low rate communications even though a reactive jammer is disrupting transmitted packets.
Actually, in \cite{wisec} two constraining assumptions are made, that is,
i) to perform an attack, the jammer first has to recognize the preamble of a packet, and ii) the jamming signal is transmitted as long  as the jammer senses activity on the channel.

In \cite{DOro2013} an analysis of energy consumption and effectiveness of a reactive jammer \textcolor{black}{attack} against timing channels is presented. 
Moreover, it is shown how a trade-off between energy consumption and jamming effectiveness \textcolor{black}{can be sought}. It is also demonstrated that 
continuous jamming can be \textcolor{black}{very} costly in terms of energy consumption.

Since the jammer and the target node(s) have opposite interests and the actions of the ones depend on those of the others, game theory is a valid tool to study such scenarios 
\cite{Wang, Yang, Altman, Sengupta}.
An anti-jamming stochastic game in cognitive radio networks is proposed in \cite{Wang}, where authors provide learning mechanisms for users to counteract jamming attacks; also,
it is shown that users can exploit frequency hopping to avoid jamming attacks by taking hopping decisions depending on the channel state.
Often the jammer has to adapt its attack depending on network operations; hence, in literature it is frequently assumed that the jammer plays as a \emph{follower} after
the \emph{leader}, i.e. the target node, has manifested its strategy. Such a scenario can be modeled as a Stackelberg game.
For example, in \cite{Yang} a Stackelberg game is proposed to model the interactions between target nodes
and a smart jammer that is able to vary its transmission power to maximize
its own utility function.
In \cite{Altman} Altman et al. analyze a game where both the target node and the jammer have energy constraints.
Finally, as specifically relevant to our work, we mention the study carried out by Sengupta et al. \cite{Sengupta} on a power control game modeling a network of nodes exploiting the timing channel,
which maximize SINR and throughput by properly setting the transmission power level and the silence duration.
In \cite{Sengupta} however, although game theory is applied to timing channel networks, jamming issues are not considered.

\textcolor{black}{As compared to the solutions proposed so far in the literature,} our paper is the first \textcolor{black}{together with \cite{anand2010attack} by Anand et al. } to develop a game-theoretical model of the interactions between 
the jammer and a target node exploiting the timing channel. \textcolor{black}{ The main differences between our work and \cite{anand2010attack} can be summarized as follows:
\begin{itemize}
	\item in \cite{anand2010attack} the target node focuses on deploying camouflaging resources (e.g., the number of auxiliary communications assisting the covert communication) to hide the underlying timing channel. In our work, instead, the target node establishes a timing channel that exploits the silence period between the end of an attack and the beginning of a subsequent packet transmission to counteract an ongoing jamming attack;
	\item in \cite{anand2010attack}, only the Nash Equilibrium (NE) is studied, whereas in our work we study both the NE and SE (Stackelberg Equilibrum). Furthermore, we compare the achievable performance of each player, and find that the SE dominates the NE (i.e., both players improve their achieved utilities), thus allowing each player to improve its own utility;
	\item in our work, the target node is able to transmit covert information even if the jammer has successfully disrupted all the bits contained in a packet. On the other hand, the authors in \cite{anand2010attack} assume that the jamming attack is successful if the Signal-to-Interference ratio (SIR) of the attacked node measured at the receiver side is higher than the one of the target node. In our approach, instead, we do not make any assumption on the SIR as, by exploiting our proposed timing channel implementation, it is possible to transmit some information even when the jammer has successfully corrupted each packet.
\end{itemize}
}

In addition, we only assume that the jammer is aware of timing channel communications ongoing between the target node and the perspective receiver, whereas
we relax the two assumptions in
\cite{wisec}. Specifically, we assume that i) to start an attack the jammer has only to detect a possible ongoing transmission activity
(e.g., the power on the monitored channel exceeds a given threshold), and ii) the transmission of the jamming signal does not necessarily
stops when the packet transmission by the target node ends, that is, the jammer is able to introduce some transmission delay in timing channel communications by extending its 
jamming signal duration.


%% file: GameGiacomo.tex
\section{Game model} \label{game}

Let us consider the scenario where two wireless nodes, a transmitter and a receiver, want to communicate, while a malicious node aims at disrupting their communication.
\textcolor{black}{To this purpose}, we assume that the malicious node executes a reactive jamming attack 
on the wireless channel.
In the following we refer to the  malicious node  as the \emph{jammer}, $J$, and the transmitting node under attack as the \emph{target node}, $T$.

\begin{figure}[t!]
\centering
   \includegraphics[width=.55\columnwidth]{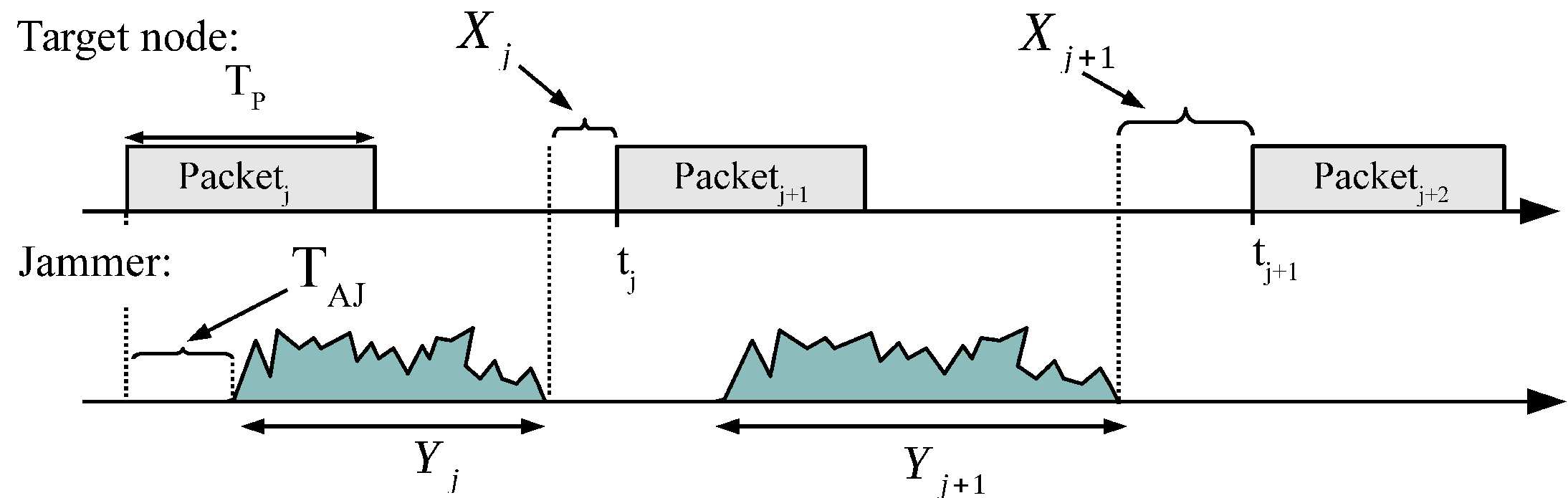}
   \caption{Interactions between the jammer and the target node.}
   \label{TransmissionScheme}
\end{figure}

The jammer senses the wireless channel continuously.
Upon detecting a possible transmission activity performed by $T$, $J$ starts emitting a jamming signal.
As shown in Fig. \ref{TransmissionScheme}, we denote as $T_{AJ}$ the duration of the time interval between the beginning of the packet transmission and
the beginning of the jamming signal emission.
The duration of the interference signal emission \textcolor{black}{that jams} the transmission of the $j$-th packet can be modeled as a continuous random variable, which we call $Y_j$.
To maximize the uncertainty on the value of $Y_j$, we assume that it is exponentially distributed with mean value $y$.

We assume that when no attack is performed the target node communicates with the receiver by applying traditional transmissions schemes; on the other hand, 
when it realizes to be under attack, it exploits the timing channel 
to transmit part of (or all) the information\footnote{Attack detection can be achieved by the target node either by means of explicit notification messages sent back to $T$ by the receiver or 
by inference after missing reception of ACK messages. Details on attack detection operations
are however out of the scope of this paper.}.
The latter is encoded in the duration of the interval between the instant when the jammer $J$ terminates the emission of the jamming signal 
and the beginning of the transmission of the next packet.
Hence, it is possible to consider a discrete time axis and refer to each timing channel utilization by means of an integer index $j$.
The silence period duration scheduled after the transmission of the $j$-th packet and the corresponding jamming signal
can be modeled as a continuous random variable, $X_j$, uniformly distributed\footnote{The uniform distribution assumption is due to the fact that, as well known, this distribution maximizes the entropy,
given the range in which the random variable is defined.}
in the range $[0, x]$.
The amount of information transmitted per each use of the timing channel depends on the value of $x$ and the precision $\Delta$ of
the clocks of the communicating nodes as shown in \cite{Morabito11}. 
\textcolor{black}{In our model we assume that the parameters $\Delta$ and $T_{AJ}$ which are hardware dependent are known a-priori to both the target node and the jammer, whereas the strategies $x$ and $y$  are estimated by means of a training phase. This is consistent with the complete information assumption which is common in game theoretic frameworks.}

To model the interactions between the target node and the jammer we propose a \textcolor{black}{jamming} game framework, defined by
a 3-tuple $\mathcal{G}=(\mathcal{N}, \mathcal{S},\mathcal{U})$, where $\mathcal{N}$ 
is the set of players, $\mathcal{S}$ is the strategy set, and $\mathcal{U}$ is the utility set.
The set $\mathcal{N}$ is composed by the target node $T$ and the jammer $J$, while the
strategy set is $\mathcal{S} =  \mathcal{S_{T}} \times  \mathcal{S_J}$, where $\mathcal{S_{T}}$ and 
$\mathcal{S_J}$ are the set of strategies of the target node and the jammer, respectively.

In our model we assume that the jammer is energy-constrained, e.g., it is battery-powered; hence, its choice of $y$
(i.e., the average duration of the signal emission that jams the packet transmission) stems from
a trade-off between two
requirements, i.e., i) reduce 
the amount of information that the target node $T$ can transmit to the perspective receiver, and ii) keep the energy consumption as low as possible.
Observe that requirement i) would result in the selection of a high value for $y$, whereas requirement ii) would result in a low value for $y$.
\textcolor{black}{On the other hand, the target node has to properly choose the value of $x$ \textcolor{black}{(i.e., the maximum silence period duration scheduled following the transmission of 
the $j-th$ packet and the subsequent jamming signal)} in order to maximize the achievable capacity $\mathcal{C}(x,y)$, i.e., the amount of information
that can be sent by means of the timing channel, while minimizing its energy consumption.}
Therefore, it is reasonable to consider that the values of $x$ and $y$ represent the \emph{strategies} for the target node $T$ and the jammer $J$, respectively.
Accordingly, the set of strategies for both players, $\mathcal{S_T}$ and $\mathcal{S_J}$, can be defined as the set of all
the feasible strategies $x$ and $y$, respectively.

The utility set of the game is defined as $\mathcal{U} = ( \mathcal{U_T} ,\mathcal{U_J})$, where $\mathcal{U_T}$ and $\mathcal{U_J}$ are
the utility functions of the target node and the jammer, respectively.
\textcolor{black}{As already said, the target node aims at maximizing its own achievable capacity, $\mathcal{C}(x,y)$ while also minimizing its energy consumption.
The jammer, on its side, aims at reducing the capacity achieved by the target node by generating interference signals, 
whose duration is $y$ (in average), while keeping its own energy consumption low.}
\textcolor{black}{Accordingly, the utility functions $\mathcal{U}_{T}(x,y)$ and $\mathcal{U}_{J}(x,y)$ to be maximized are defined as follows:
\begin{equation} 
{
      \left \{
      \begin{array}{rcl}
       \mathcal{U}_{T}(x,y)&=&+C(x,y)-c_{T^*} \cdot T_P  \cdot P_T \\ \label{utility_scm}
       \mathcal{U}_{J}(x,y)&=&-C(x,y)-c_T \cdot y  \cdot P
      \end{array}
      \right.
}
\end{equation}
\noindent
where $P_T$ and $P_J$ are the transmission power of the target node and the jammer, respectively, $T_P$
is the duration of a transmitted
packet in seconds, $c_{T^*}$ and $c_{T}$ are positive transmission costs expressed in [bit/(s $\cdot$ J)] which weight
the two contributions in the utility functions and therefore, in the following will be referred to as \emph{weight parameters}.
\textcolor{black}{Note that while the energy consumption of the jammer varies as a function of the strategy $y$ of the jammer itself,
on the contrary the energy consumption of the target node during a cycle only depends on the duration $T_P$ of the packet and not on the strategy.}}
Furthermore, a low value of $c_T$ means that the jammer considers its jamming effectiveness more important than its energy consumption, 
while a high $c_T$ value indicates that the jammer is 
energy-constrained and, as a consequence, it prefers to save energy rather than reducing the capacity of the target node.
We observe that $c_T=0$ models the case of continuous jamming without any energy constraint, which is of limited interested and out of the scope of this paper, 
since we focus on studying the trade-off between the achievable capacity and the consumed energy.


Let us now calculate the capacity $C(x,y)$ which appears in the utility function (\ref{utility_scm}).
To this purpose, we denote the interval between two consecutive transmissions executed by $T$ as a \emph{cycle}.
The expected duration of a cycle is
\begin{equation} \label{TCycleCalculation}
	t_{\mathrm{Cycle}}=T_{AJ} + y + x/2
\end{equation}
\noindent
The capacity $C(x,y)$ can be derived as the expected value of the information transferred during a cycle, 
$c_{\mathrm{Cycle}}(x,y)$, divided by the \textcolor{black}{expected} duration of
a cycle, $t_{{\mathrm{Cycle}}}$.
It is easy to show that $c_{{\mathrm{Cycle}}}(x,y)$ is approximately 
\begin{equation}
	c_{{\mathrm{Cycle}}} = \log_2\left(x/\Delta \right) 
	\label{capacityBasic}
\end{equation}
\noindent
Note that at each timing channel utilization the target node $T$ is expected to transmit at least one bit; then, from eq. (\ref{capacityBasic}) it follows that $x \geq 2 \Delta$.

Eqs. (\ref{TCycleCalculation}) and (\ref{capacityBasic}) can be exploited to calculate the capacity $C(x,y)$, i.e.,
\begin{equation} \label{Cxy2}
	C(x,y)= 
	\frac{
		\log_2\left(
		x /\Delta \right) 
	}{
		T_{AJ} + y + x/2
	}
\end{equation}

\begin{minipage}[t]{0.4\textwidth}
\centering
   \includegraphics[width=\columnwidth]{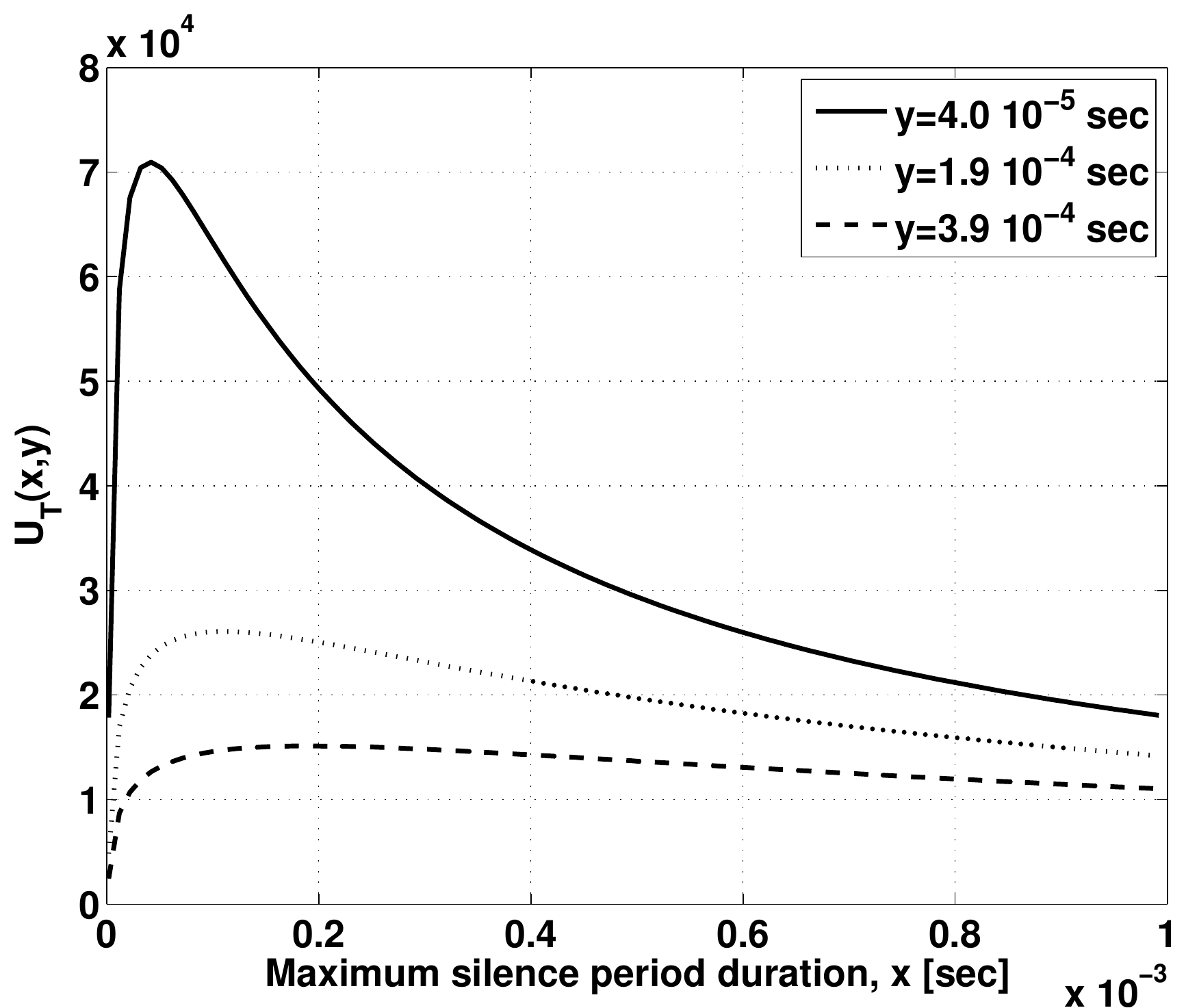}
   \captionof{figure}{Utility function of the target node ($\mathcal{U}_{{T}}(x,y)$ ) as a function of $x$ for different values of the average jamming signal duration $y$ ($c_{T^*} \cdot P = 2 \cdot 10^6$).}
   \label{uTC}
\end{minipage} \hspace{0.1\textwidth}
\begin{minipage}[t]{0.4\textwidth}
\centering
   \includegraphics[width=\columnwidth]{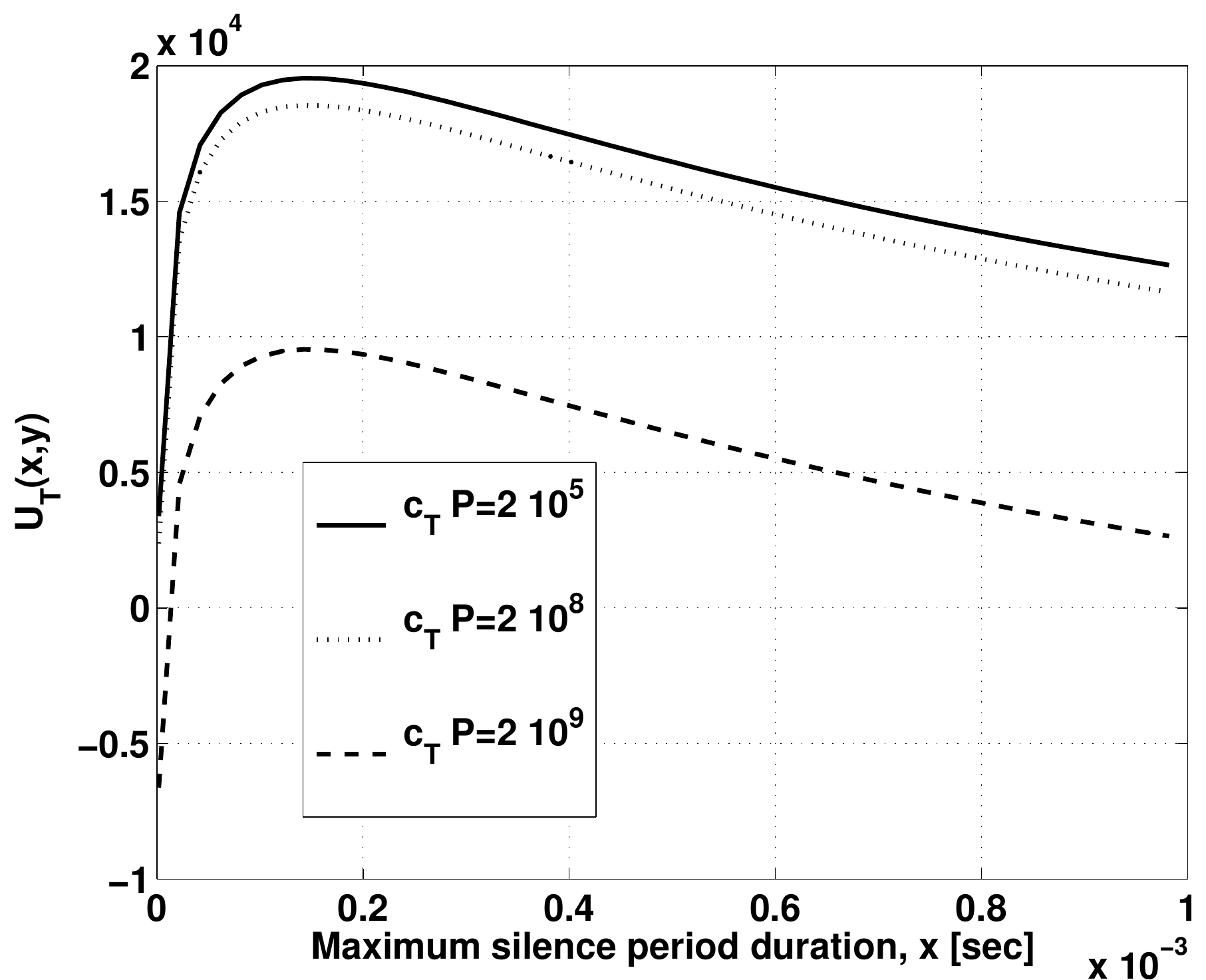}
   \captionof{figure}{Utility function of the target node ($\mathcal{U}_{{T}}(x,y)$ ) as a function of $x$ for different values of the product $c_{T^*} \cdot P$ ($y=2.8 \cdot 10^{-4} s$).}
   \label{uTC_CT}
\end{minipage} \vspace{1cm}

Hereafter we illustrate a simple numerical example that refers to the same realistic scenarios addressed in \cite{Wilhelm}.
The considered parameter settings are reported in Table \ref{SimulationParameters}.
It is also assumed that \textcolor{black}{both the target node and the jammer transmit their respective signals by using the maximum allowed transmitting power, i.e., $P_T=P_J=P$.} 
\begin{table}[t]
  \begin{center} 
    \begin{tabular}{|l|l|l|}
	    \hline
	    Name & Value & Unit\\
	    \hline
	    $T_{AJ}$ & 15 & $\mu$s \\
	    \hline
	    $\Delta$ & 1 & $\mu$s \\
	    \hline
	    $P$ & 2 & W \\
	    \hline
        $T_{P}$ & 50 & $\mu$s \\
	    \hline
    \end{tabular}
    \end{center}
    \caption{\label{SimulationParameters} Parameter settings used in our simulations.}
\end{table}

Fig. \ref{uTC} shows the utility function of the target node $T$ as a function of $x$, for different values of $y$. 

We note that $\mathcal{U}_{{T}}(x,y)$
increases when $x$ increases  until it reaches a threshold after which the utility function starts decreasing.
This is due to the fact that, when $x$ is higher than such a threshold, the silence duration is large enough 
to cause an increase in the transmission delay and, consequently, a decrease in the transmission capacity.
This is a well known result in timing channel communications \cite{Morabito07}.
In Fig. \ref{uTC} we also note that the achievable performance noticeably depends on the jamming signal duration $y$.
In fact, when $y$ increases,
the capacity of the target node decreases a
s the jamming attack forces the transmitter in delaying its timing channel communications by increasing $x$.
\textcolor{black}{Figure \ref{uTC_CT} shows the impact of the energy consumption on the utility achieved by the target node. As expected, the higher the product $c_{T} \cdot P$ is, the lower the achieved utility is. Note that, as the energy consumption in any cycle is constant and does not depend on either $x$ or $y$,  the energy cost of the target node $\mathcal{U}_{{T}}(x,y)$ would only result in a slight shift in the utility function of the target node.}

Fig. \ref{uJ} shows instead the utility function of the jammer $\mathcal{U}_{J}(x,y)$ vs. $y$ for different values of $x$. 

\begin{minipage}[t]{0.4\textwidth}
\centering
   \includegraphics[width=\columnwidth]{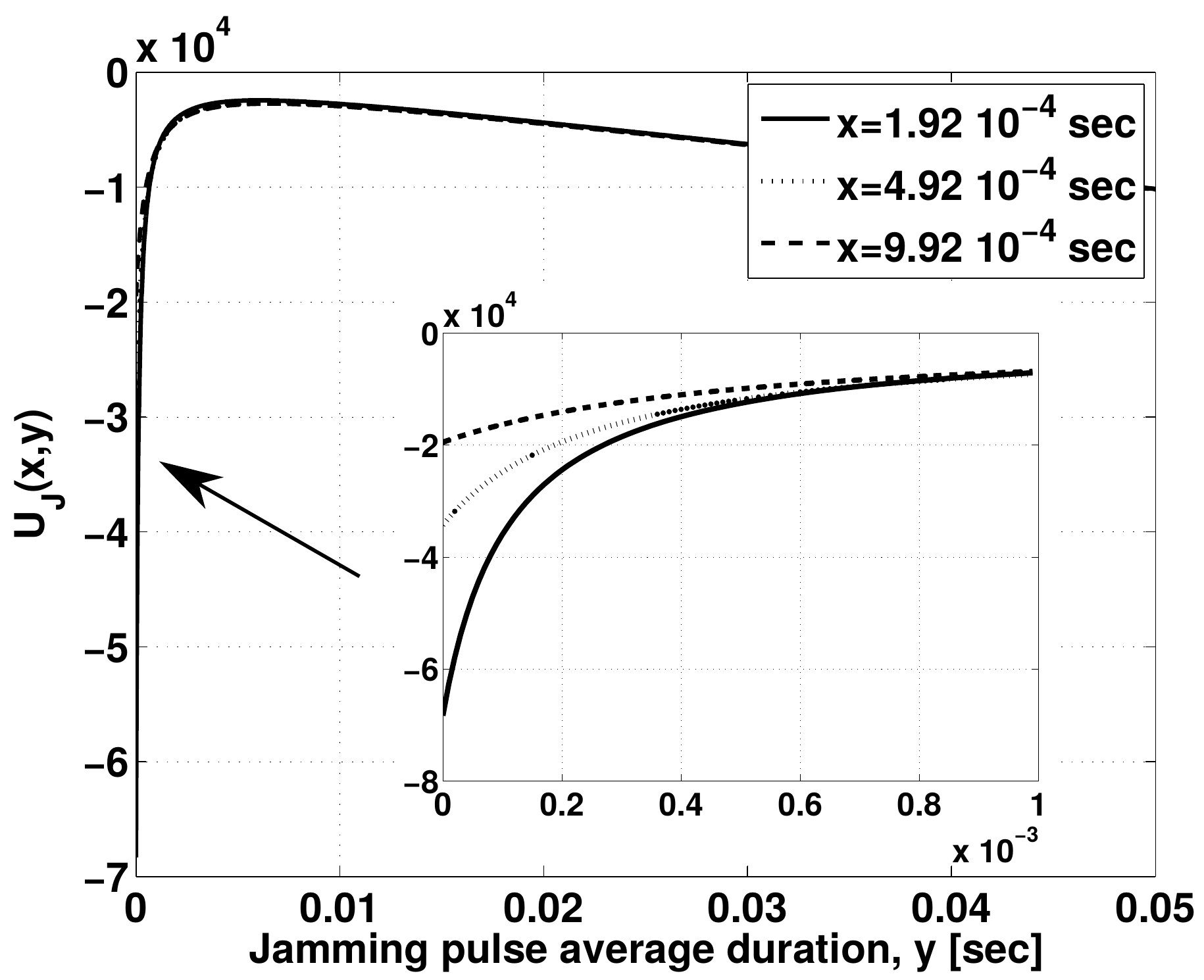}
   \captionof{figure}{Utility function of the jammer ($\mathcal{U}_{J}(x,y)$) as a function of $y$ for different values of the maximum silence period duration $x$ ($c_T \cdot P = 2 \cdot 10^6$).}
   \label{uJ}
\end{minipage}\hspace{0.1\textwidth}
\begin{minipage}[t]{0.4\textwidth}
\centering
 \includegraphics[width=\columnwidth]{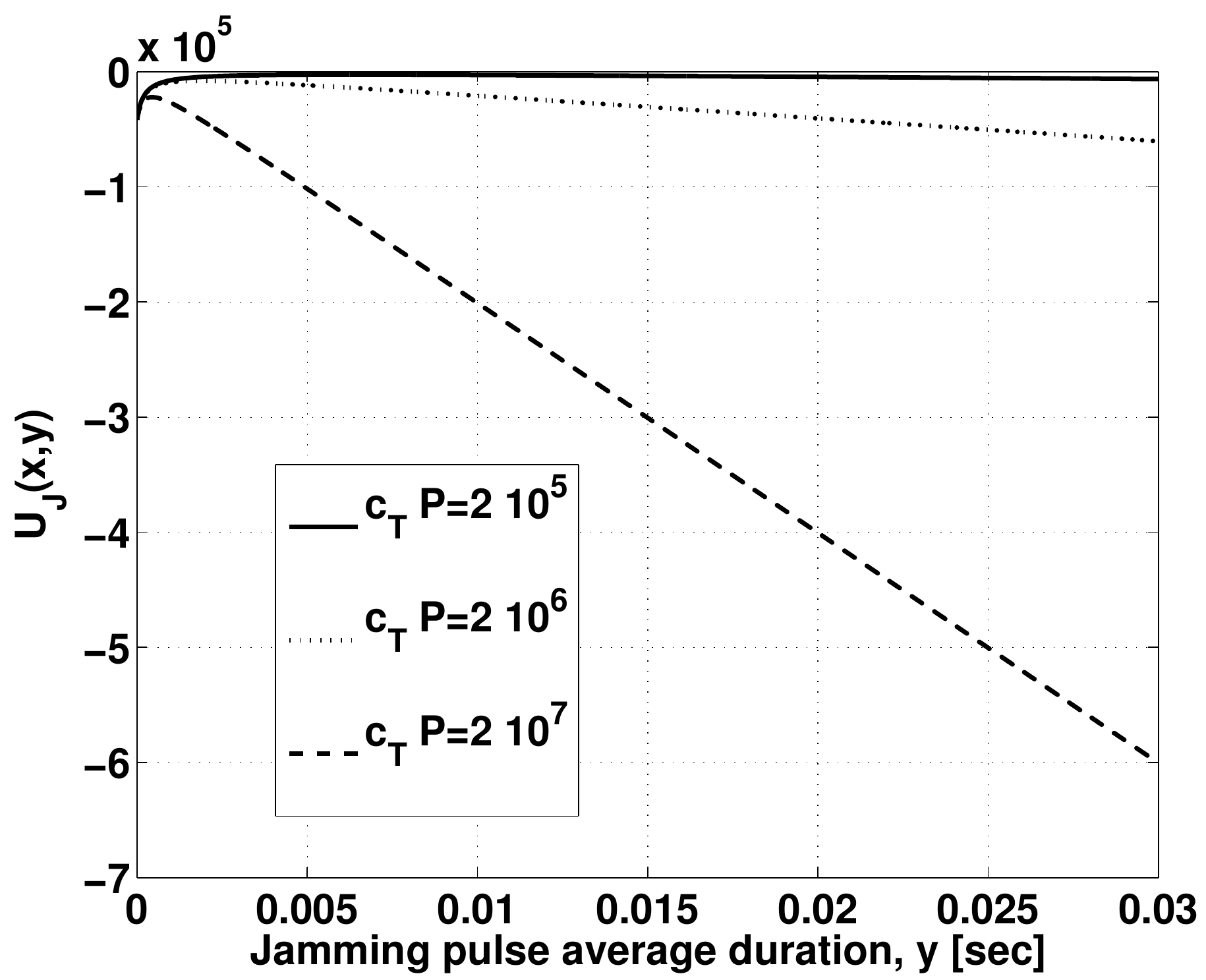}
   \captionof{figure}{Utility function of the jammer ($\mathcal{U}_{J}(x,y)$) as a function of $y$ for different values of the product $c_T \cdot P$ ($x=5 \cdot 10^{-4} s$).}
   \label{uJvscT} 
\end{minipage} \vspace{1cm}

Note that for high values of $y$ the utility function $\mathcal{U}_{J}(x,y)$ does not practically depend on $x$.
This is because high $y$ values imply $C(x,y)\approx0$ regardless of the specific value of $x$. Such a behavior is evident in Fig. \ref{uJ}.
We observe that for high values of $y$ the capacity achieved by the target node $C(x,y)$ is negligible
and, thus, the utility function of the jammer can be approximated as $\mathcal{U}_{J}(x,y) \simeq - c_T \cdot y \cdot P$. In other words, the utility of the jammer decreases
linearly with $y$.
\textcolor{black}{To this purpose}, in Fig. \ref{uJvscT} we show the utility of the jammer $\mathcal{U}_{{J}}(x,y)$ for different values of the product $c_T \cdot P$. 
It is \textcolor{black}{evident that, as expected,} when the cost of transmitting the interference signal
\textcolor{black}{at the jammer} is high (i.e., $c_T \cdot P$ is high) the utility function $\mathcal{U}_{{J}}(x,y)$ decreases rapidly \textcolor{black}{and linearly}. 
%

%% file: NE_SCMGiacomo3_short_appendix.tex
\section{Nash Equilibrium analysis} \label{NE_SCM}

In this Section we solve the game described in Section \ref{game}, and
we find the \emph{Nash Equilibrium points} (NEs), in which both players achieve their highest utility given
the strategy profile of the opponent. 
In the following we also provide proofs of the {\it existence, uniqueness} and {\it convergence} to the Nash Equilibrium under best response dynamics.

Let us recall the definition of Nash equilibrium:
\begin{definition}
 A strategy profile $(x^*,y^*) \in \mathcal{S}$ is a  Nash Equilibrium (NE) if $ \forall (x',y') \in \mathcal{S}$
 \begin{align}
       \mathcal{U}_{T}(x^*,y^*) \geqslant \mathcal{U}_{T}(x',y^*) \nonumber \\
       \mathcal{U}_{J}(x^*,y^*) \geqslant \mathcal{U}_{J}(x^*,y') \nonumber
 \end{align}
 that is, $(x^*,y^*)$ is a strategy profile where no player has incentive to deviate unilaterally.
\end{definition}

One possible way to study the NE and its properties is to look at the \emph{best response functions} (BRs).
A best response function is a function that maximizes the utility function of a player, given the opponents' strategy profile. 
Let $b_{T}(y)$ be the BR of the target node and $b_J(x)$ the BR of the jammer. These functions can be characterized as follows:

\begin{equation}
 b_{T}(y)  = \arg\max_{x \in \mathcal{S_T}} \mathcal{U_T}(x,y) \label{btbef} \nonumber
\end{equation}
\begin{equation}
  b_J(x)  = \arg\max_{y \in \mathcal{S_J}} \mathcal{U_J}(x,y) \label{bjbef} \nonumber
\end{equation}
\noindent
In our model it is possible to analytically derive the closed form of the above BRs by analyzing the first derivatives of $\mathcal{U_T}(x,y)$ and $\mathcal{U_J}(x,y)$, 
and imposing that $\frac{\partial}{\partial x}\mathcal{U_T}(x,y) = 0$ and $\frac{\partial}{\partial y}\mathcal{U_J}(x,y) = 0$.

It is easy to see that $\frac{\partial}{\partial x}\mathcal{U_T}(x,y) = 0$ leads to 
\begin{equation}
 \frac{1}{x}-\frac{1}{2} \log\left(\frac{x}{\Delta}\right)\frac{1}{T_{AJ}+y+\frac{x}{2}}=0 \label{deriveBT}
\end{equation}
\noindent
Eq. (\ref{deriveBT}) can be rewritten as follows:
\begin{equation}
\frac{2(T_{AJ}+y)}{e \Delta} = \frac{x}{e\Delta} \cdot \log{{\frac{x}{e\Delta}} }\label{deriveBT2}
\end{equation}
Note that eq. (\ref{deriveBT2}) is in the form $\beta=\alpha \log\alpha$, and, by exploiting the definition of \emph{Lambert W-function}, say $W(z)$,
which, for any complex $z$, satisfies $z=W(z)e^{W(z)}$, it has solution $\alpha = e^{W(\beta)}$.

Therefore, eq. (\ref{deriveBT2}) can also be rewritten as
\begin{equation}
 x=\Delta e^{W\left( \frac{2(T_{AJ}+y)}{e\Delta}\right)+1} \label{deriveBTfinal} \nonumber
\end{equation}
which is, by definition, $b_T(y)$.

In order to derive the closed form of $b_J(x)$ we first solve $\frac{\partial}{\partial y}\mathcal{U_J}(x,y) = 0$.
It can be easily proven that $\frac{\partial}{\partial y}\mathcal{U_J}(x,y) = 0$ leads to
\begin{equation}
 \log\left(\frac{x}{\Delta}\right)= \eta \left(T_{AJ}+\frac{x}{2}+y\right)^2 \nonumber
\end{equation}
which can be rewritten as follows:
\begin{equation}
 b_J(x)=\sqrt{\frac{\log(\frac{x}{\Delta})}{\eta}}-T_{AJ}-\frac{x}{2}  \label{deriveBJfinal} \nonumber
\end{equation}
where $\eta = c_T\cdot  P \cdot \log2$.

Therefore, we can write
\begin{equation}
 b_{T}(y)  =  \Delta e^{ \psi(y) +1} \label{bestTC} 
\end{equation}
\begin{equation}
  b_J(x)  = 
	  \begin{cases}
	    \chi(x) , & \mbox{if } \chi(x) \geqslant 0 \\ 
	    0 , & \mbox{if } \chi(x) < 0 
	  \end{cases}
	  \label{bestJ} 
\end{equation}
\noindent
where
\begin{equation}
 \psi(y) = W\left( \frac{2[T_{AJ}+y]}{e \Delta} \right) \hspace{1cm}  \hspace{1cm}
 \chi(x) = \sqrt{\frac{\log(\frac{x}{\Delta})}{\eta}} - T_{AJ} - \frac{x}{2}
 \label{ChiFunction}
\end{equation}
\noindent
Note that the best response of the jammer $b_J(x)$ depends on the value of the weight parameter $c_T$. Also, it can be shown that there exists a critical value of the
weight parameter, say $c_T^{(\mathrm{max})}$, such that $b_J(x)<0$ $\forall x \in \mathcal{S_T}$, $\forall c_T \geq c_T^{(\mathrm{max})}$. 
In fact, since the function $\chi(x)$ is strictly decreasing in $c_T$, $\lim_{c_T \rightarrow +\infty} \chi(x) < 0$ and $\lim_{c_T \rightarrow 0} \chi(x) = +\infty$,
the \emph{intermediate value theorem} ensures the existence of $c_T^{(\mathrm{max})}$. By looking at the first derivative of the $\chi(x)$ function in eq. (\ref{ChiFunction}), it can be shown
that $c_T^{(\mathrm{max})}=\frac{1}{P \log(2)}\frac{1}{2\Delta (\Delta+T)}$.
Therefore, if $c_T \geq c_T^{(\mathrm{max})}$ the only possible strategy of the jammer is $b_J(x)=0$, and then, as the strategy 
set of the jammer ($\mathcal{S_J}$) is a singleton, the game has a trivial outcome.

\subsection{Existence of the Nash Equilibrium}

It is well known that the intersection points between $b_{T}(y)$ and $b_{J}(x)$ are the NEs of the game.
Therefore, to demonstrate the existence  of at least one NE, it suffices to prove that
$b_{T}(y)$ and $b_{J}(x)$ have one or more intersection points.
In other words, it is sufficient to find one or more pairs $(x^*,y^*) \in \mathcal{S}$ such that 
\begin{equation}
 (b_{T}(y^*),b_{J}(x^*))=(x^*,y^*)
 \label{bestrespInter}
\end{equation}
\noindent

To this aim, in the following we provide some structural properties of the utility functions, $\mathcal{U_T}(x,y)$ and $\mathcal{U_J}(x,y)$,
that will be useful in solving eq. (\ref{bestrespInter}).

\begin{lemma}
\label{lemma:utility_property}
For the utility functions $\mathcal{U_T}(x,y)$ and $\mathcal{U_J}(x,y)$, the following properties hold \footnote{The proof of Lemma \ref{lemma:utility_property} 
which is straightforward (although quite long),
consists in calculating  the first and second derivatives of the utility functions and studying them.
}:
\begin{itemize}
\item $\mathcal{U_T}(x,y)$ is strictly concave for $x \in [2\Delta,x']$ and is monotonically decreasing for $x>x'$ where $x' = b_T(y)$
\item $\mathcal{U_J}(x,y)$ is strictly concave  $\forall y \in \mathcal{S_J}$.
\end{itemize} 
\end{lemma}

\begin{theorem}[NE existence]
\label{theorem:NEexist}
The game $\mathcal{G}$ admits at least an NE.
\end{theorem}

\begin{IEEEproof}
If we limit the strategy of the target node to $[2\Delta,x']$, it follows from Lemma~\ref{lemma:utility_property} that there exists at least an NE 
since both the utility functions are concave in the restraint strategy set \cite{Rosen}. 
However, this does not still prove the existence of the NE in the non-restraint strategy set $\mathcal{S_T}$.
Let $(x^*,y^*)$ denote the NE with a restraint strategy set $[2 \Delta, x']$; we can easily observe that $(x^*,y^*)$ is also the NE of the jamming game
with non-restraint strategy set. 
To show this, recall Lemma~\ref{lemma:utility_property} that states that $\mathcal{U_T}(x,y)$ is monotonically decreasing for $x>x'$. 
The transmitter has thus no incentive to deviate from $(x^*,y^*)$ and the jammer has no incentive to deviate from it either. 
Therefore, $(x^*,y^*)$ is the NE of the jamming game.
\end{IEEEproof}

\subsection{Uniqueness of the Nash Equilibrium}

After proving the NE existence in Theorem \ref{theorem:NEexist}, let us prove the uniqueness of the NE, that is, there is only one strategy
profile such that no player has incentive to deviate unilaterally.

\begin{theorem}[NE uniqueness]
 The game $\mathcal{G}$ admits a unique NE that can be expressed as
 \begin{equation}
  (x_{\mathrm{NE}},y_{\mathrm{NE}})= 
  \begin{cases}
   \left(\Delta e^{\frac{1}{2}W(\frac{8}{\eta \Delta^2})} , \frac{\Delta}{2}[\frac{1}{2}W(\frac{8}{\eta \Delta^2})-1]e^{\frac{1}{2}W(\frac{8}{\eta \Delta^2})}-T_{AJ}\right) & \mbox{if } c_T < \tilde{c_T} \\
   \left(\Delta e^{W(\frac{2T}{e\Delta})+1} , 0\right) & \mbox{otherwise}
   \label{NEequilibriumCLosed}
  \end{cases}
 \end{equation}
where  $\eta=c_T\cdot  P \cdot \log2$ and
 \begin{equation}
  \tilde{c_T}=\frac{4}{\Delta^2P\log2} e^{-2[W(\frac{2T}{e\Delta})+1]}/(W(\frac{2T}{e\Delta})+1)
  \label{cTTilde}
 \end{equation}
 \label{theorem:NEunique}
\end{theorem}

	\textcolor{black}{The proof consists in exploiting formal and structural properties of the best response functions to show that their intersection is unique,
	that is, eq. (\ref{bestrespInter}) admits a unique solution. For a detailed proof see \ref{app:NEunique}}

\subsection{Convergence to the Nash Equilibrium} \label{sec:convergence}

We now analyze the convergence of the game to the NE when players follow \emph{Best Response Dynamics} (BRD).
In BRD the game starts from any initial point $(x^{(0)},y^{(0)}) \in \mathcal{S}$ and, at each successive step, each player plays its strategy by following
its best response function.
Thereby, at the $i$-th iteration the strategy profile $(x^{(i)},y^{(i)})$ can be formally expressed by the following BRD iterative algorithm:
\begin{equation}
 \begin{cases}
  x^{(i)}=b_{T}(y^{(i-1)}) \\
  y^{(i)}=b_J(x^{(i-1)})
 \end{cases}
 \label{iterative}
 \nonumber
\end{equation}

Let $\textbf{b}(x,y)=(b_T(y),b_J(x))^T$ be the best response vector and $J_\textbf{b}$ be the Jacobian of $\textbf{b}(x,y)$ defined as follows
\begin{equation}
 J_\textbf{b}=\begin{bmatrix}
       \frac{\partial}{\partial x} b_T(y) & \frac{\partial}{\partial y} b_T(y) \\[0.8em]
       \frac{\partial}{\partial x} b_J(x)& \frac{\partial}{\partial y} b_J(x)
\end{bmatrix} = 
\begin{bmatrix}
       0 & \frac{\partial}{\partial y} b_T(y) \\[0.3em]
       \frac{\partial}{\partial x} b_J(x) & 0
\end{bmatrix}
\label{jacobian}
\end{equation}
\noindent
\textcolor{black}{It has been demonstrated \cite{luenberger1978complete} that, if the Jacobian infinity matrix norm
$|| J_\textbf{b} ||_{\infty} < 1$, 
the BRD always converges to the unique NE.}
In the following we prove the following theorem:

\begin{theorem}[NE convergence - sufficient condition]
The relationship 
\begin{equation} \label{cTCondition}
	c_T > \frac{1}{9 \Delta^2 \log2 P} \frac{1}{\left(W(\frac{2T_{AJ}}{e\Delta})+1\right) e^{2(W(\frac{2T_{AJ}}{e\Delta})+1)}}
\end{equation}
\noindent
is a sufficient condition for the game $\mathcal{G}$ to converge to the NE. \textcolor{black}{Furthermore, it converges to the NE in at most $\log_{J^{max}_b}\frac{\epsilon}{||s^1-s^0||}$ iterations for any $\epsilon$, where $J^{max}_b=\max J_\textbf{b}$ and $s^i=(x^i,y^i)$.}
\end{theorem}

To demonstrate the theorem, 
\begin{enumerate}
	\item
		\textcolor{black}{we prove that the relationship
		\begin{equation} \label{Lemma2Conclusion}
			\max_{x \in \mathcal{S_T}} \left( \frac{1}{\eta x^2\log(\frac{x}{\Delta})} \right) < 9
		\end{equation}
		\noindent
		is a sufficient condition for the BRD to converge to the NE \textcolor{black}{in at most $\log_{J^{max}_b}\frac{\epsilon}{||s^1-s^0||}$ iterations}.
		This is the focus of Lemma \ref{LemmaConditionOnBothXandCt}};
	\item
		we define a game $\mathcal{G}'$ and demonstrate that $\mathcal{G}$ converges to $\mathcal{G}'$ in two iterations at most.
		This is the focus of Lemma \ref{lemmaFromGPrimeToG};
	\item
		we demonstrate that the condition in eq. (\ref{cTCondition}) is a sufficient condition for $\mathcal{G}'$ to satisfy eq. (\ref{Lemma2Conclusion}) and converge to the same NE of $\mathcal{G}$. 
		This is the focus of Lemma \ref{lemmaFromGPrimeToNash}.
\end{enumerate}


\begin{lemma} \label{LemmaConditionOnBothXandCt}
 \textcolor{black}{ The BRD converges to the unique NE from any $(x^{(0)},y^{(0)}) \in \mathcal{S}$ 
 if $\max_{x \in \mathcal{S_T}} \left( \frac{1}{\eta x^2\log(\frac{x}{\Delta})} \right) < 9$ in at most $\log_{J^{max}_b}\frac{\epsilon}{||s^1-s^0||}$ iterations}.
   \label{lemma:convergenceLarge}
\end{lemma}

	\textcolor{black}{The proof is based on showing that the above relationship is a sufficient condition
	for the Jacobian infinity matrix norm $|| J_\textbf{b} ||_{\infty}$ to be always lower than 1, and thus, according to \cite{luenberger1978complete}, convergence of the BRD follows. 
	We refer the reader to \ref{app:lemmaConvergence} for a detailed proof of Lemma 2.}

Let us now observe that $b_J(x)$ is lower-bounded as it is non-negative ($ b_J (x) \geqslant 0$) and, since it is concave, it has a maximum, say $y_M$, for 
$\hat{x}=\Delta e^{\frac{1}{2}W(\frac{2}{\eta \Delta^2})}$, and thus it is upper-bounded ($ b_J (x) \leqslant y_M=b_J(\hat{x}) $).
Also,  it is easy to prove that $b_T(y)$ is a non-negative strictly increasing function, hence, it is lower-bounded by $x_m=b_T(0)$.
We can thus define a new strategy set $\mathcal{S}' = \mathcal{S_T}' \times \mathcal{S_J}' = [x_m,x_M] \times [0,y_M]$,
where $\mathcal{S'} \subset \mathcal{S}$ and $x_M=b_T(y_M)$, which is relevant in the following lemma:
%
%
\begin{lemma} \label{lemmaFromGPrimeToG}
 Given any starting point $(x^{(0)},y^{(0)}) \in \mathcal{S}$, the BRD is bounded in $\mathcal{S'}$ in at most two iterations.
 That is, $(x^{(i)},y^{(i)}) \in \mathcal{S}'$ for $i=2,3,...,+\infty$.
 \label{lemma:limitedBRD}
\end{lemma}
\begin{IEEEproof}
 Let $\mathcal{S}^{(1)} $ be the strategy set at the first iteration.
 From eqs. (\ref{bestTC}) and (\ref{bestJ}) we have that
 $b_J(x)$ is lower and upper-bounded by $y=0$ and $y=y_M$, respectively, thus $y^{(1)} \in [0,y_M]$. 
 Furthermore, as $b_T(x)$ is lower-bounded by $x=x_m$ and $y^{(0)} \in \mathcal{S_J} = [0, +\infty [$, it follows that $x^{(1)} \in [x_m,+ \infty)$.
 Hence, we have that 
 $\mathcal{S}^{(1)}=\mathcal{S_T}^{(1)} \times \mathcal{S_J}^{(1)}= [x_m,+ \infty) \times [0,y_M ]$, $\mathcal{S}^{(1)} \subset \mathcal{S} $.
 Due to the boundedness of $y^{(1)}$ which assumes values in $\mathcal{S_J}^{(1)}$, it can be shown that at the second iteration 
 $x^{(2)} \in [x_m,x_M]$ while $y^{(2)} \in [0,y_M]$, thus, we have that $(x^{(2)},y^{(2)}) \in \mathcal{S}'$.
 We can extend the same reasoning to the $j$-th iteration ($\forall j=3,4,...,\infty$) to obtain that $(x^{(j-1)},y^{(j-1)}) \in \mathcal{S}'$. Therefore,
 it follows that $(x^{(j)},y^{(j)})$ 
 is still in $\mathcal{S}'$, which concludes the proof.
\end{IEEEproof}


\begin{lemma}[NE convergence] \label{lemmaFromGPrimeToNash}
If the parameter $c_T$ satisfies the condition:
\begin{equation} \label{ConditionOnCT}
 c_T >c_T'= \frac{1}{9 \Delta^2 \log2 P} \frac{1}{\left(W(\frac{2T_{AJ}}{e\Delta})+1\right) e^{2(W(\frac{2T_{AJ}}{e\Delta})+1)}} 
\end{equation}
\noindent
then $\mathcal{G}'$ converges to the NE of $\mathcal{G}$.
\end{lemma}
\begin{IEEEproof}
Since the function on the left-hand side of eq. (\ref{Lemma2Conclusion}) is non-negative and strictly decreasing, and the minimum value of $\mathcal{S_T}$ is $x_m = \Delta e^{ W\left(\frac{2T_{AJ}}{e\Delta} \right) +1}$, then 
\begin{equation} \label{maximumOfTheEq}
	\max_{x \in \mathcal{S_T}} \left( \frac{1}{\eta x^2\log(\frac{x}{\Delta})} \right) = \frac{1}{\eta x_m^2\log(\frac{x_m}{\Delta})}
\end{equation}

\textcolor{black}{It is easy to show that if eq. (\ref{ConditionOnCT}) holds, then  }
\begin{equation}
	 \frac{1}{\eta x_m^2\log(\frac{x_m}{\Delta})} < 9
	 \nonumber
\end{equation}
\noindent
and therefore, recalling eq. (\ref{maximumOfTheEq}), eq. (\ref{Lemma2Conclusion}) holds.
From Lemma \ref{LemmaConditionOnBothXandCt} we thus obtain that $\mathcal{G}'$ converges to its NE.

We still need to demonstrate that $\mathcal{G}$ and $\mathcal{G}'$ converge to the same equilibrium point.
To this purpose it is sufficient to prove that the equilibrium point of $\mathcal{G}$ is in $\mathcal{S}'$.
Theorem \ref{theorem:NEunique} guarantees that the game $\mathcal{G}$ admits a unique equilibrium, which has to be in $\mathcal{S}$. 
Let $(x_{\mathrm{NE}},y_{\mathrm{NE}})$ be the NE, i.e., the unique intersection point between $b_T(y)$ and $b_J(x)$.
As $b_J(x)$ takes values in $[0,y_M]$ it follows that $y_{\mathrm{NE}} \in [0,y_M]$; therefore, $x_{\mathrm{NE}} = b_T(y_{\mathrm{NE}}) \in [x_m,x_M]$. 
It follows that $(x_{\mathrm{NE}},y_{\mathrm{NE}}) \in \mathcal{S}'$, which concludes the proof.
\end{IEEEproof}


\section{Stackelberg Game} \label{SE_SCM}

In a Stackelberg game one of the players acts as the leader by anticipating the best response of the follower. 
In our scenario, the jammer plays its strategy when a communication from the target node is detected on the monitored channel;
thus, it is natural to assume that the target node acts as the leader followed by the jammer.
Obviously, given the strategy of the target node $x$, the jammer will play the strategy that maximizes its utility, that is, its best response $b_J(x)$\footnote{\textcolor{black}{In the following, given that the value of $c_{T^*}$ does not impact on the game, for worth of simplicity we assume that $c_{T^*}=0$. }}.
This hierarchical structure of the game allows the leader to achieve a utility which 
is at least equal to the utility achieved in the ordinary game $\mathcal{G}$ at the NE,
if we assume \emph{perfect knowledge}, that is, the target node is completely aware of the utility function of the jammer and its 
parameters, and thus it is able to evaluate $b_{J}(x)$. 
Whereas, if some parameters in the utility function of the jammer are unknown at the target node, i.e., the \emph{imperfect knowledge} case, the above result is no more 
guaranteed as it is impossible to evaluate the exact form of $b_{J}(x)$.
In this section we analyze the Stackelberg game and provide useful results about its equilibrium points, referred to as Stackelberg Equilibria (SEs).

\begin{definition}
 A strategy profile $(x^*,y^*) \in \mathcal{S}$ is a Stackelberg Equilibrium (SE) if $y^* \in \mathcal{S_J}^{\mathrm{NE}}(x)$ and
 \begin{align}
       x^*=\arg \max_{x'} \mbox{ } \mathcal{U_T}(x',y^*)
       \nonumber
 \end{align}
 where $\mathcal{S_J}^{\mathrm{NE}}(x)$ is the set of NE for the follower when the leader plays its strategy $x$.
\end{definition}

In the following we will prove that, in the case of perfect knowledge, there is a unique SE for any value of the weight parameter $c_T$, and
we demonstrate that the target node can inhibit the jammer under the perfect knowledge assumption. 
Next, we will investigate the implications of imperfect knowledge on the game outcome.

\subsection{Perfect Knowledge} \label{sec:perfect}
Under the perfect knowledge assumption, the target node selects $x$ in such a way that $\mathcal{U}_{T}(x,b_{J}(x))$ is maximized, where $\mathcal{U}_{T}(x,b_{J}(x))$ can be 
calculated by replacing expression (\ref{bestJ}) in eqs. (\ref{Cxy2}) and (\ref{utility_scm}) as follows
%

\begin{subnumcases}{\mathcal{U}_{T}(x,b_{J}(x))=}
  \sqrt{c_T P \log_2(\frac{x}{\Delta})}-c_{T^*} \cdot T_P \cdot P_T  \hspace{1.0cm} \mbox{  if } \chi(x)>0 \label{SEut1} \\ 
  \log_2(\frac{x}{\Delta})/(T_{AJ}+\frac{x}{2}) -c_{T^*} \cdot T_P \cdot P_T  \hspace{0.6cm}  \mbox{  if } \chi(x) \leqslant 0 \label{SEut2}
\end{subnumcases}
\noindent
%
By analyzing the first derivative of $\chi(x)$, it can be shown that $\chi(x)$ has a maximum in $\hat{x}=\Delta e^{\frac{1}{2}W(\frac{2}{\eta \Delta^2})}$
and, consequently, $\chi(x)$ is strictly decreasing for $x>\hat{x}$ and strictly increasing for $x<\hat{x}$. 

In the following we show that for any value of $c_T$ there exists a unique Stackelberg Equilibrium,
and this is when the jammer does not jam the timing channel\footnote{In this case the jammer is expected to transmit the interference signal
for a short time interval only because this suffices to disrupt communications, as occurs in traditional communication channels.}.
Furthermore, we show that the leader can improve its utility at the Stackelberg equilibrium if and only if $c_T<\tilde{c_T}$.

\begin{figure}[t]
\centering
   \includegraphics[scale=0.55]{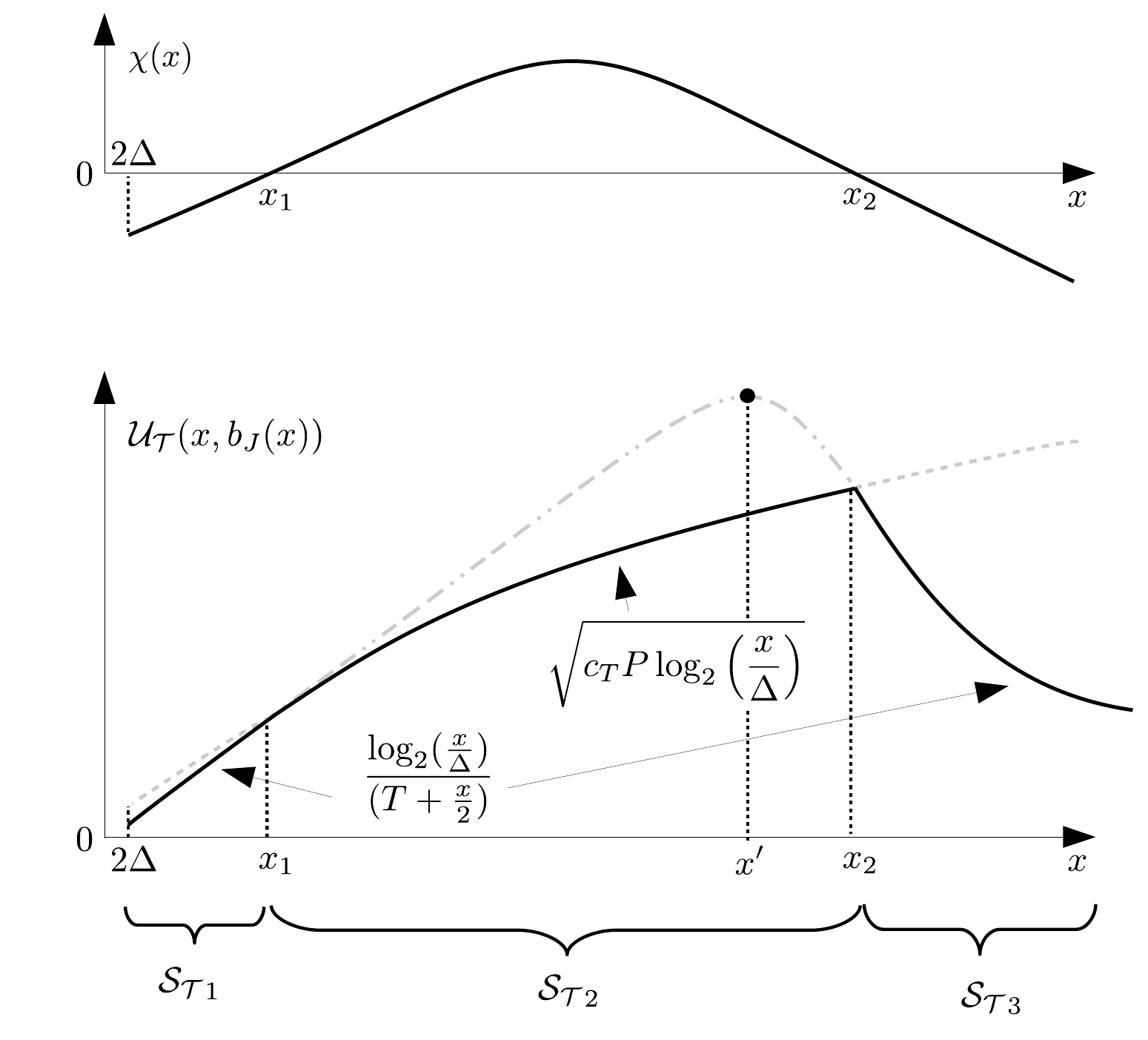}
   \caption{Graphical representation of $\chi(x)$ and $\mathcal{U_T}(x,b_J(x))$ in the Stackelberg game. The solid line is the actual utility of the target node in each strategy subset.}
   \label{graphicChi}
\end{figure}

\begin{theorem}
 For any value of the parameter $c_T$, the Stackelberg game $\mathcal{G_T}$ has a unique equilibrium. 
 \label{theorem:SET}
\end{theorem}
\begin{IEEEproof}
    	First, we prove that the game admits a unique equilibrium for $c_T \geq c_T^{(\mathrm{max})}$.
 Recall that $c_T \geq c_T^{(\mathrm{max})}$ implies $b_J(x)=0$; therefore, $\mathcal{S_J}$
 is singleton and the unique feasible strategy for the jammer at the SE is $y_{\mathrm{SE}}=0$. 
 In fact, due to the high cost associated to the emission of the jamming signal, 
 the jammer is inhibited $\forall x \in \mathcal{S_T}$. Hence, it can be easily proved that the strategy
 profile at the SE is $(x_{\mathrm{SE}},y_{\mathrm{SE}})=(\Delta e^{W(\frac{2T_{AJ}}{e\Delta})+1},0)$, that is, at the SE the target node selects the strategy
 that maximizes the capacity of the non-jammed timing channel (where indeed $y_{\mathrm{SE}}=0$).
 
 Instead, if $c_T < c_T^{(\mathrm{max})}$, from eq. (\ref{ChiFunction}) we have that $\chi(\hat{x})>0$. Thus, for the intermediate value theorem 
 there exist $x_1<\hat{x}$ and $x_2>\hat{x}$ such that $\chi(x_1)=\chi(x_2)=0$, as shown in Fig. \ref{graphicChi}. 
 
 Let us denote $\mathcal{S_T}_1=\{x \in [2\Delta,x_1]\}$, $\mathcal{S_T}_2=\{x \in [x_1,x_2]\}$,
 $\mathcal{S_T}_3=\{\mathcal{S_T} \smallsetminus (\mathcal{S_T}_1 \cup \mathcal{S_T}_2) \}$, and $x'=\Delta e^{W(\frac{2T_{AJ}}{e\Delta})+1}$.
 It can be easily proved that $x'$ maximizes eq. (\ref{SEut2}) and, since $\chi(x')>0$, it follows that $x' \in \mathcal{S_T}_2$.
 Therefore, the utility function of the target node as defined in eq. (\ref{SEut2}) increases for $x<x'$ and decreases for $x>x'$.
 The latter is fundamental to prove the theorem; in fact, as shown in Fig. \ref{graphicChi}, for $x \in \mathcal{S_T}_1$ the utility of the target node 
 is defined by eq. (\ref{SEut2}) and strictly increases as $x$ increases; therefore, we have that in $\mathcal{S_T}_1$ the maximum utility is achieved
 in $x_1$. On the contrary, in $\mathcal{S_T}_2$ the utility is defined by eq. (\ref{SEut1}), which is a strictly increasing function that achieves its maximum value for
 $x=x_2$. Finally, for $x \in \mathcal{S_T}_3$ we have that the utility of the transmitter defined by eq. (\ref{SEut2}) strictly decreases as $x>x'$; hence,
 the maximum value is achieved for $x=x_2$.
 
 Since $\mathcal{U}_{T}(x,b_{J}(x))<\mathcal{U}_{T}(x_2,b_{J}(x_2))$ with $x \neq x_2$, it follows that,
 to maximize its own utility, the target node must play the unique strategy $x=x_2$.
 Note that $\chi(x_2)=0$ by definition, thus from eq. (\ref{bestJ}) we have that the strategy of the jammer at the equilibrium is $y_{\mathrm{SE}}=0$.
 Therefore, $x_{\mathrm{SE}}=x_2$ is the strategy of the target node at the SE, and we can 
 identify the unique SE as $(x_{\mathrm{SE}},y_{\mathrm{SE}})=(x_2,0)$, which concludes the proof.
\end{IEEEproof}
 
 Let us remark that the above Theorem also highlights an insightful side-effect: at the Stackelberg equilibrium, pursuing the goal of inhibiting the jammer makes the target node
 prefer to increase transmission delay rather than reduce its achievable capacity.
 
\textcolor{black}{Let us also note that, although an analytical closed form for $x_{\mathrm{SE}}$ cannot be easily derived, its value can be determined by means of numerical search algorithms such as the bisection search algorithm.
Obviously, such algorithms will not give the exact value of $x_{\mathrm{SE}}$; in fact, they will return an interval $[x_m, x_M]$ small as desired, containing the solution, i.e.,  $x_{\mathrm{SE}} \in [x_m, x_M]$, and eventually the target node will select the minimum or the maximum value of the interval which gives the highest utility function.
Let $\epsilon(x_m, x_M)$ denote the loss in the utility of the target node due to the fact that it cannot determine the exact value of $x_{\mathrm{SE}}$.
Given that the utility function is continuous and that its derivative is upperbounded by $u_{\max}= \sqrt{c_T \cdot P}/(4 \Delta \log 2)$ in $[x_1, x_{\mathrm{SE}}]$, it is possible to show that selecting the interval size in such a way that
\begin{equation}
        x_M-x_m \leq \epsilon^*/u_{\max}
\end{equation}
\noindent
the loss in the utility of the target node, $\epsilon(x_m, x_M)$, is lower than $\epsilon^*$.
In other terms, by using numerical search algorithms such as the bisection search algorithm, the target node can make the loss in its utility as small as desired.}

In the following we provide an approximation $x_{\mathrm{SE}}'$ that can be helpful from a practical point of view.
Let us assume that $\left( T_{AJ}+\frac{x}{2}\right) \approx \frac{x}{2}$, therefore, eq. (\ref{ChiFunction}) can be rewritten as follows
\begin{equation}
 \frac{\log(\frac{x}{\Delta})}{\log(2) c_T P} = \left(\frac{x}{2}\right)^2
 \label{eq:xTildeApprox}
\end{equation}
By means of simple \textcolor{black}{manipulations} it can be easily shown that eq. (\ref{eq:xTildeApprox}) admits the following solution:
\begin{equation}
 x_{\mathrm{SE}}'=\Delta e^{-\frac{1}{2}W\left(-\frac{\log(2)c_TP\Delta^2}{2}\right)}
 \label{eq:xTildeApproxFinal}
\end{equation}
In Section \ref{numericalres} we will provide numerical results that show how much the approximation in eq. (\ref{eq:xTildeApproxFinal}) affects
the outcome of the Stackelberg game.

\begin{theorem}
In the Stackelberg game the target node improves its utility as compared to the NE if and only if $0<c_T<\tilde{c_T}$.
 \label{theorem:improvement}
\end{theorem}

\begin{IEEEproof}
%
Let us start with the proof of the sufficient condition implied by the Theorem \ref{theorem:improvement}.
According to eqs. (\ref{NEequilibriumCLosed}) and (\ref{SEut1}), proving that
$\mathcal{U}_{T}(x_{\mathrm{SE}},b_{J}(x_{\mathrm{SE}})) > \mathcal{U}_{T}(x_{\mathrm{NE}},y_{\mathrm{NE}})$ is equivalent to showing that 
 \begin{equation}
   \sqrt{c_T P \log_2(\frac{x_{\mathrm{SE}}}{\Delta})} >\frac{1}{\log 2} \frac{2}{\Delta}e^{-\frac{1}{2}W(\frac{8}{\eta \Delta^2})}
   \nonumber
 \end{equation}
 that is
 \begin{equation}
  \frac{1}{2} W(\frac{8}{\eta \Delta^2}) < \log(\frac{x_{\mathrm{SE}}}{\Delta})
  \nonumber
 \end{equation}
 This only holds if $x_{\mathrm{SE}} > \Delta e^{\frac{1}{2}W(\frac{8}{\eta \Delta^2})} = x_{\mathrm{NE}}$.
 Recall that if $0 < c_T < \tilde{c}_T$, the NE is an interior NE, that is, $\chi(x_{\mathrm{NE}})>0$.
 Therefore, as $\chi(x_{\mathrm{SE}})=0$, it must hold that $x_{\mathrm{NE}}<x_{\mathrm{SE}}$, which proves the sufficiency condition.
As for the necessary condition, we have to show that,
if $c_T\geqslant\tilde{c_T}$, no improvement can be achieved by the target node. 
In fact, \textcolor{black}{if $c_T\geqslant\tilde{c_T}$} it is straightforward to prove that the NE and the SE coincide, and thus, 
the utilities of the target node at the SE and NE are equal.
\end{IEEEproof}

\subsection{Imperfect knowledge} \label{sec:imperfect}

We now investigate the implications of \emph{imperfect knowledge} on the weight parameter $c_T$ in eq. (\ref{utility_scm}).
In Theorem \ref{theorem:SET} we proved that the optimal strategy in the Stackelberg game is $x_{\mathrm{SE}}$ such that $\chi(x_{\mathrm{SE}})=0$.
According to eq. (\ref{ChiFunction}) the value of $c_T$ is needed to evaluate $x_{\mathrm{SE}}$.
However, it is reasonable to assume that in realistic scenarios the value of $c_T$ is not available at the target node, while instead, 
only statistical information on the distribution
of $c_T$ is likely known.
Let us denote  as $f_{c_T}(\xi)$ the probability density function (pdf) of the random variable representing the weight parameter $c_T$.
We also denote as $g(\xi)$ the function returning the strategy of the target node at the SE, $x_{\mathrm{SE}}$, when the weight parameter for the jammer is $c_T=\xi$.

The resulting utility function of the target node ${\bf U}_{T}^{\xi}=\mathcal{U}_{T}(g(\xi),b_J(g(\xi))$ can be calculated as
\begin{subnumcases}{{\bf U}_{T}^{\xi}=}
			\sqrt{c_T P \log_2\left(\frac{g(\xi)}{\Delta}\right)}  \hspace{1.8cm} \mbox{  if } \xi > c_T \label{imperfect_u_t1} \\ 
			\log_2\left(\frac{g(\xi)}{\Delta}\right)/\left(T_{AJ}+\frac{g(\xi)}{2}\right) \hspace{0.6cm}  \mbox{  if } \xi \leq c_T \label{imperfect_u_t2}
\end{subnumcases}
\noindent
Let us refer to $E\{{\bf U}_{T}^{\xi}\}$ as the expected value of the utility function of the target node. 
Assuming that $f_{c_T}(\xi)$ is \textcolor{black}{a continuous function}, it follows that 
\begin{equation}
\begin{array}{l}
E\{{\bf U}_{T}^{\xi}\}=\int_{-\infty}^{+\infty}\mathcal{U}_{T}(\xi|c_T=\alpha)f_{c_T}(\alpha)d\alpha=\int_{-\infty}^{\xi}\mathcal{U}_{T}(\xi|c_T=\alpha)f_{c_T}(\alpha)d\alpha+\int_{\xi}^{+\infty}\mathcal{U}_{T}(\xi|c_T=\alpha)f_{c_T}(\alpha)d\alpha
\end{array}
\nonumber
\end{equation}
From eqs. (\ref{imperfect_u_t1}) and (\ref{imperfect_u_t2}) we have
\begin{eqnarray}
 E\{{\bf U}_{T}^{\xi}\}=\int_{-\infty}^{\xi}\sqrt{\alpha P \log_2(\frac{g(\xi)}{\Delta})}f_{c_T}(\alpha)d\alpha+\int_{\xi}^{+\infty}\frac{\log_2(\frac{g(\xi)}{\Delta})}{(T_{AJ}+\frac{g(\xi)}{2})}f_{c_T}(\alpha)d\alpha = \nonumber \\
 = \sqrt{P \log_2(\frac{g(\xi)}{\Delta})}\int_{-\infty}^{\xi}\sqrt{\alpha}f_{c_T}(\alpha)d\alpha+\frac{\log_2(\frac{g(\xi)}{\Delta})}{(T_{AJ}+\frac{g(\xi)}{2})}\int_{\xi}^{+\infty}f_{c_T}(\alpha)d\alpha
 \label{eq:imperfectIntermediate}
\end{eqnarray}
\noindent
By exploiting the relationship in eq. (\ref{ChiFunction}), eq. (\ref{eq:imperfectIntermediate}) can be rewritten as
\begin{equation}
 E\{{\bf U}_{T}^{\xi}\}= P \left(T_{AJ}+\frac{g(\xi)}{2}\right)\sqrt{\xi}\left[\int_{-\infty}^{\xi}\sqrt{\alpha}f_{c_T}(\alpha)d\alpha+\sqrt{\xi}\int_{\xi}^{+\infty}f_{c_T}(\alpha)d\alpha\right]
 \label{eq:expectedGeneral}
\end{equation}

Note that the target node has first to find $\xi^*=\arg\max_{\xi}E\{{\bf U}_{T}^{\xi}\}$, and then, the optimal strategy is evaluated as $x_{\mathrm{SE}}\left(\xi^*\right)$ such that
$\chi\left(x_{\mathrm{SE}}(\xi^*)\right)=0$.

In the following we analyze the especially relevant case when the random variable $\xi$ is uniformly 
distributed in a closed interval\footnote{Note that the uniform distribution represents the worst case, as it is the distribution that maximizes the uncertainty on the actual value of $c_T$, given that a minimum and a maximum values are given.},
that is, the pdf of $\xi$ is defined as
\begin{equation}
f_{c_T}(\xi)=
 \begin{cases}
			\frac{1}{\xi_{max}-\xi_{min}}  \hspace{0.6cm} \mbox{  if } \xi \in [\xi_{min},\xi_{max}] \\ 
			0  \hspace{0.6cm} \mbox{  otherwise }
\end{cases}
\label{eq:cTdistro}
\end{equation}
\noindent
By substituting eq. (\ref{eq:cTdistro}) in eq. (\ref{eq:expectedGeneral}), we obtain the following expression
\begin{equation}
 E\{{\bf U}_{T}^{\xi}\}=P \frac{\left(T_{AJ}+\frac{g(\xi)}{2}\right)}{\xi_{max}-\xi_{min}}\left[ \xi\xi_{max}-\frac{1}{3}\xi^2-\frac{2}{3}\xi^\frac{1}{2}\xi_{min}^{\frac{3}{2}} \right]
 \label{eq:expectedUniform}
\end{equation}
\noindent
In order to maximize the expected utility we study the first derivative of eq. (\ref{eq:expectedUniform}), which leads to:
\begin{equation}
 \frac{W\left(-\frac{P\log(2)\Delta^2}{2}\xi\right)}{1+W\left(-\frac{P\log(2)\Delta^2}{2}\xi\right)}
		      \left( \xi_{max}-\frac{1}{3}\xi-\frac{2}{3}\frac{\xi_{min}^\frac{3}{2}}{\sqrt{\xi}} \right) =
		      2\xi_{max}-\frac{4}{3}\xi-\frac{2}{3}\frac{\xi_{min}^\frac{3}{2}}{\sqrt{\xi}}
 \label{eq:expectedFirstDerivativeGeneral}
\end{equation}

The solution of eq. (\ref{eq:expectedFirstDerivativeGeneral}), say $\xi_{opt}$, is the value of $\xi$ that maximizes the expected utility of the target node.
Regrettably, $\xi_{opt}$ can be evaluated only numerically.
Thus, \textcolor{black}{in the aim of } providing practical methods to choose $\xi$, 
in the next \textcolor{black}{section} we will \textcolor{black}{discuss} some analytical 
results that show how $\xi=\xi_{max}$ well approximates $\xi_{opt}$.
In fact, if we assume $W\left(-\frac{P\log(2)\Delta^2}{2}\xi\right)/\left[1+W\left(-\frac{P\log(2)\Delta^2}{2}\xi\right)\right] \approx 1$, then, eq. (\ref{eq:expectedFirstDerivativeGeneral}) 
can be reformulated as 
\begin{equation}
 \xi_{max}-\frac{1}{3}\xi=2\xi_{max}-\frac{4}{3}\xi
 \label{eq:expectedFirstDerivativeApprox}
 \nonumber
\end{equation}
\noindent
whose solution is $\xi=\xi_{max}$.
Furthermore, we will show that the above approximation guarantees 
high efficiency at the SE even if the uncertainty on the actual value of $c_T$ is high, as in the case of a uniform distribution.

%% file: Numerical.tex
\section{Numerical Results} \label{numericalres}


In this section we apply the theoretical framework developed in the previous sections to numerically analyze the equilibrium properties for both the ordinary 
and Stackelberg games.
As introduced in Section \ref{game}, the settings of the relevant parameters are those in Table \ref{SimulationParameters}.

\subsection{Ordinary Game} \label{sec:ordinary}

\begin{figure}[htb!]
\centering
   \includegraphics[scale=0.55]{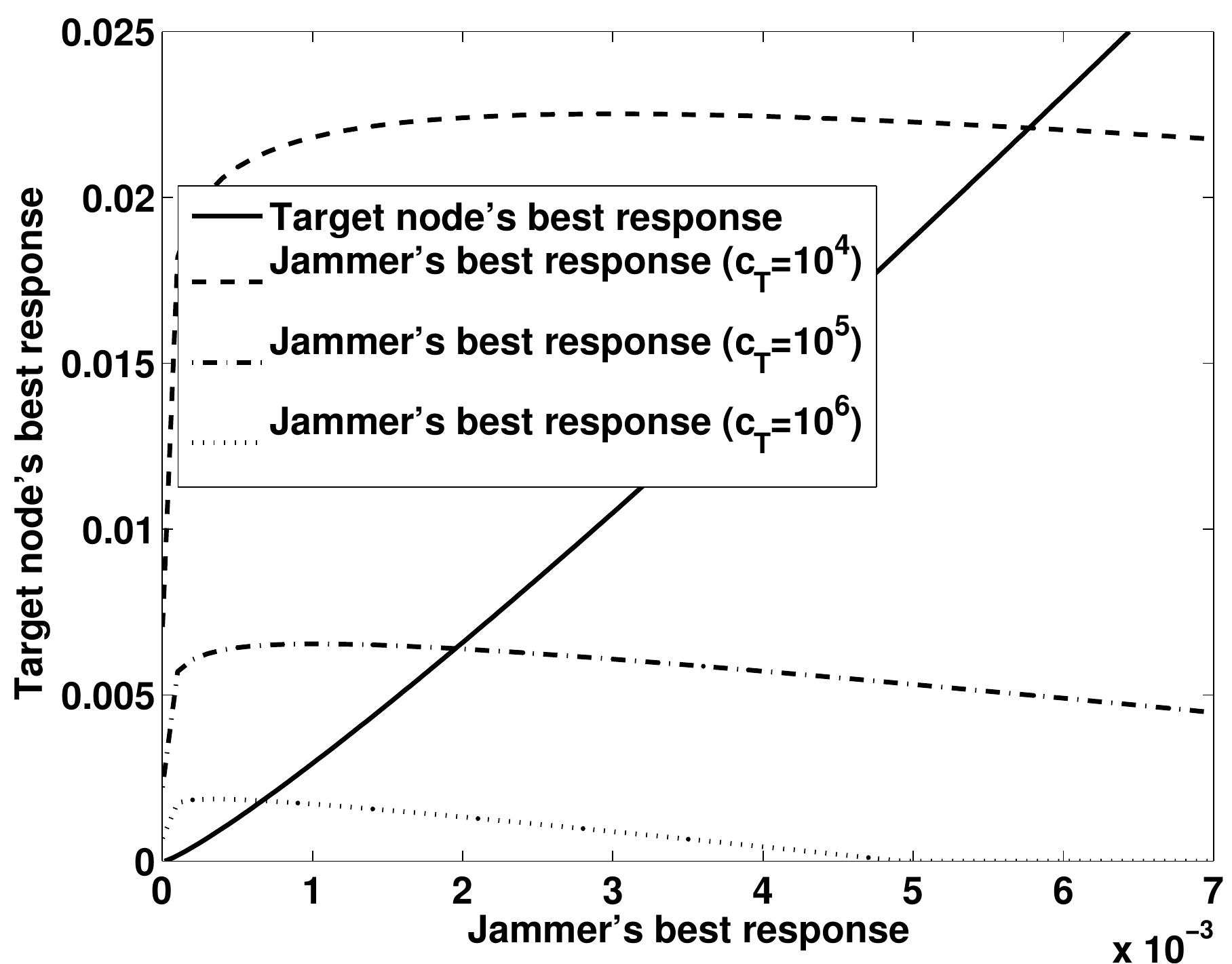}
   \caption{Best response functions for both the target node and the jammer.}
   \label{intersectionBRs}
\end{figure}

In Fig. \ref{intersectionBRs} we show the best response functions of both the target node and the jammer for different values of the weight \textcolor{black}{parameter} $c_T$.
As already said, the NE is the intersection point between the best response functions. As expected, the best response of the target node 
does not depend on the value of $c_T$, while this is not true for 
the best response of the jammer. Note that for high $c_T$ values the jammer reduces its jamming signal duration $y$, and the strategy of the target node consists in reducing
the maximum silence duration $x$.

\begin{figure}
\centerline{
\subfigure[]{
\includegraphics[width=.45\textwidth]{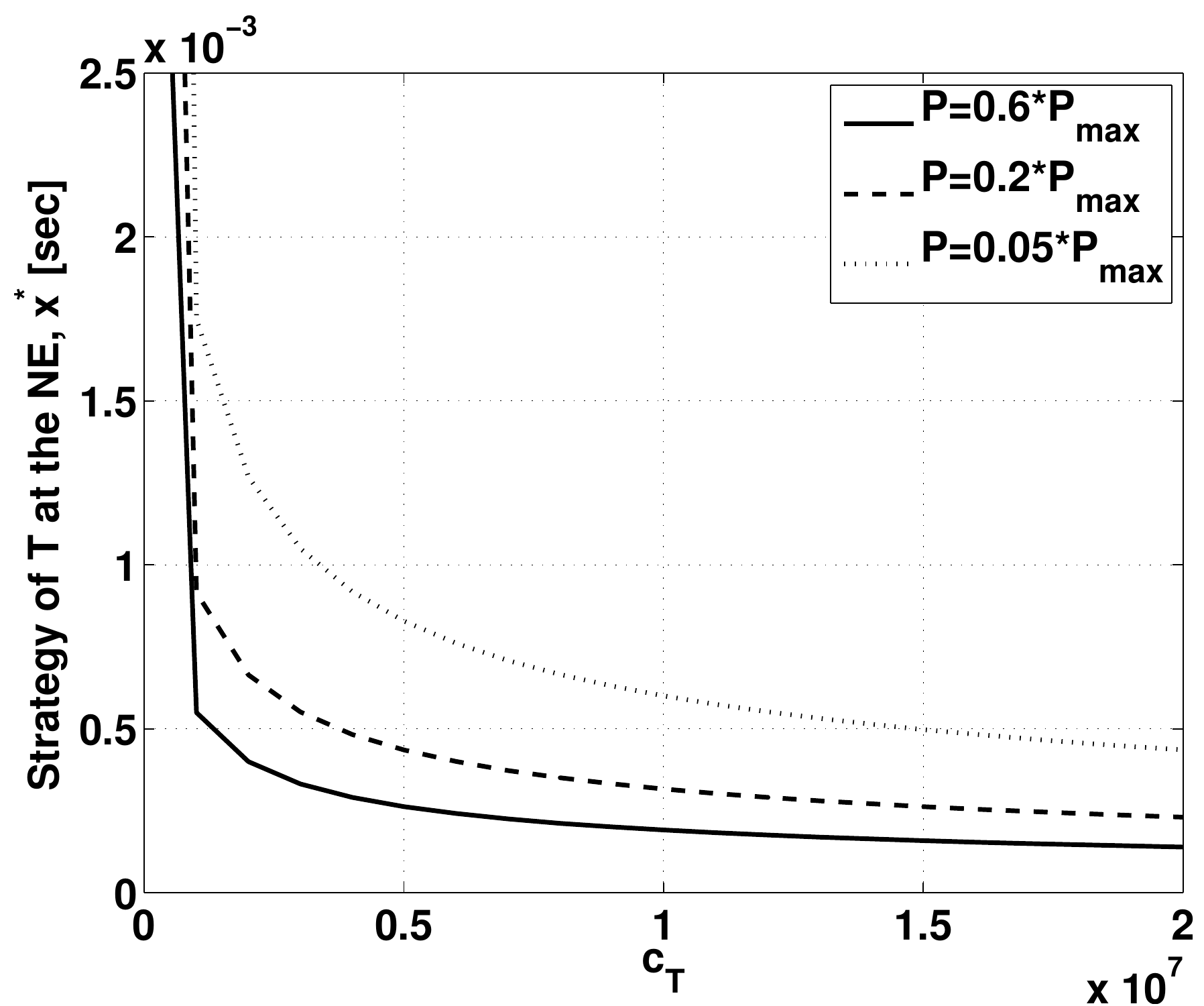}\label{loopNEx}
}
\hfil
\subfigure[]{
\includegraphics[width=.45\textwidth]{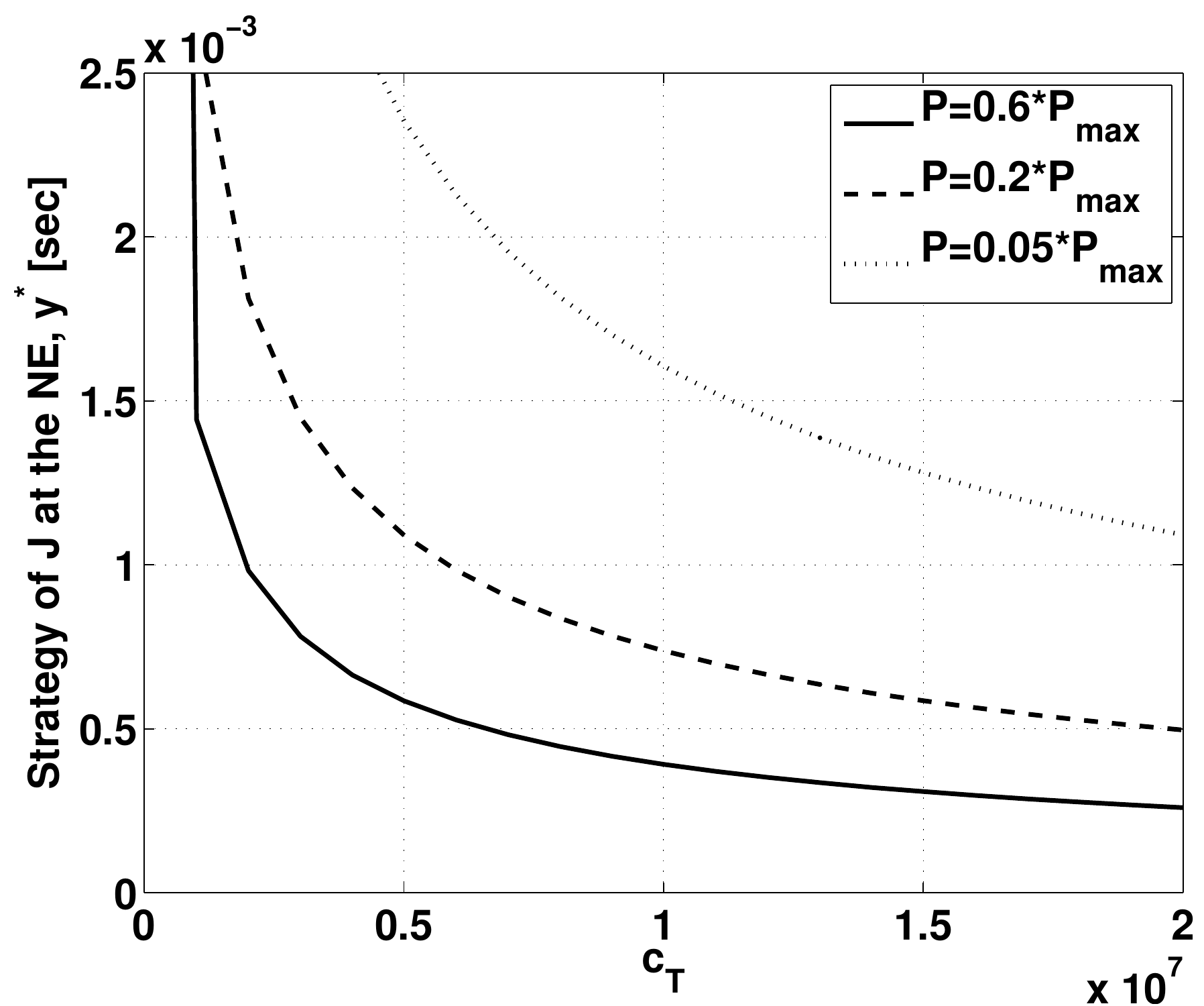}\label{loopNEy}
}
}
\caption{a) Strategy of the target node at the NE as a function of the weight parameter $c_T$ for different values of the transmitting power $P$ b) Strategy of the jammer at the NE as a function of the weight parameter $c_T$ for different values of the transmitting power $P$.}
\label{}
\end{figure}

Figs. \ref{loopNEx} and \ref{loopNEy} illustrate the strategy of the players at the NE as a function of $c_T$ for different values of the transmitting power $P$. 
Note that, as $c_T$ increases, the target node decreases the maximum silence duration and the jammer reduces the jamming signal duration as well.
In fact, upon increasing $c_T$ the jammer acts in an energy preserving fashion and this causes a decrease in
$y$. Such a behavior allows the target node to behave more aggressively by reducing the maximum silence duration $x$.
Furthermore, upon increasing $P$, the strategies $x$ and $y$ decrease as higher $P$ values force the jammer to reduce the jamming signal duration and, thus,  the energy
consumption. Also, the target node can reduce $x$, thus increasing its achieved capacity. 

\begin{figure}
\centerline{
\subfigure[]{
\includegraphics[width=.45\textwidth]{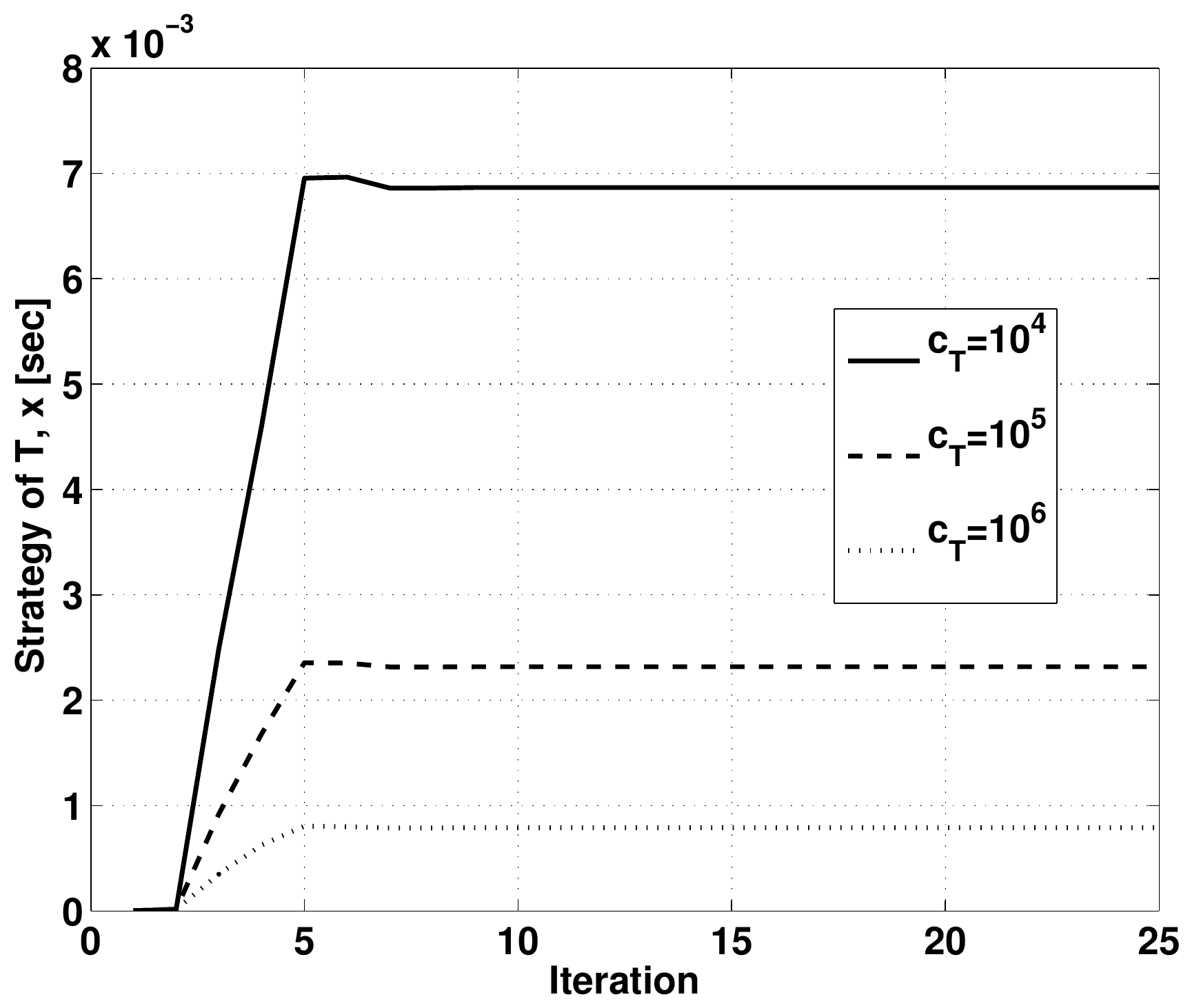}\label{iterX}
}
\hfil
\subfigure[]{
\includegraphics[width=.45\textwidth]{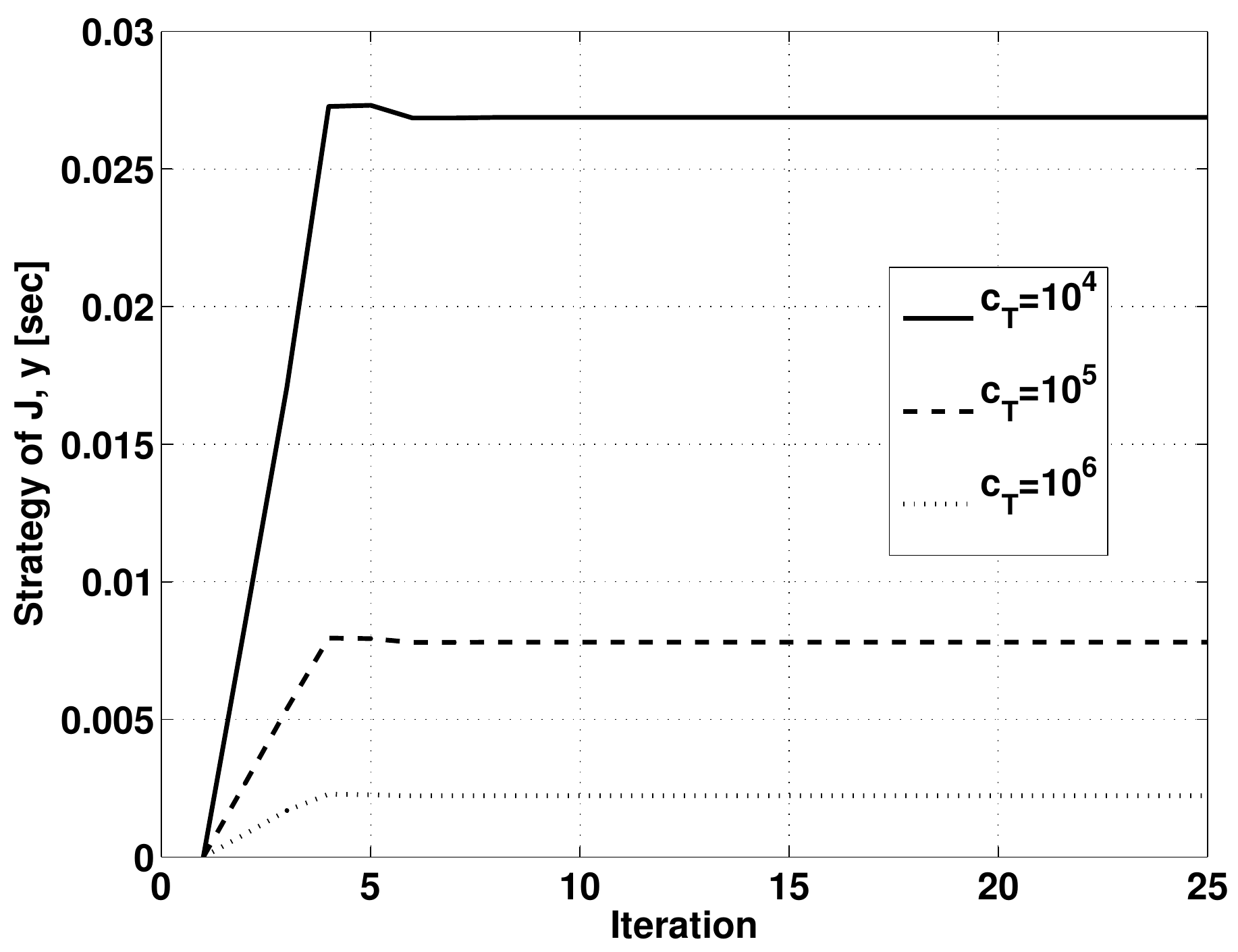}\label{iterY}
}
}
\caption{a) Strategy of the target node at each iteration b) Strategy of the jammer at each iteration.}
\label{}
\end{figure}

Figs. \ref{iterX} and \ref{iterY} illustrate how the BRD evolves at each iteration for different values of the weight $c_T$. Since we proved that the game converges
to the NE, Figs. \ref{iterX} and \ref{iterY} show how, as expected, the players' strategies converge
\textcolor{black}{to the strategy set $\mathcal{S'}$ in 2 iterations (as discussed in Lemma \ref{lemmaFromGPrimeToG}) and} to the NE in at 
most 7 iterations\footnote{Note that, although we proved that the convergence to the NE is guaranteed only if $c_T<\tilde{c_T}$, in our simulations the game always converges to the NE in a few iterations, independently of the value of $c_T$.}.
\textcolor{black}{It is also shown that an increase in the value of $c_T$ causes a decrease in the strategies of both players due to the aggressive behavior of the jammer.}

\subsection{Stackelberg Game} \label{sec:stackelberg}
%
%
\begin{figure}[t]
\centering
   \includegraphics[scale=0.5]{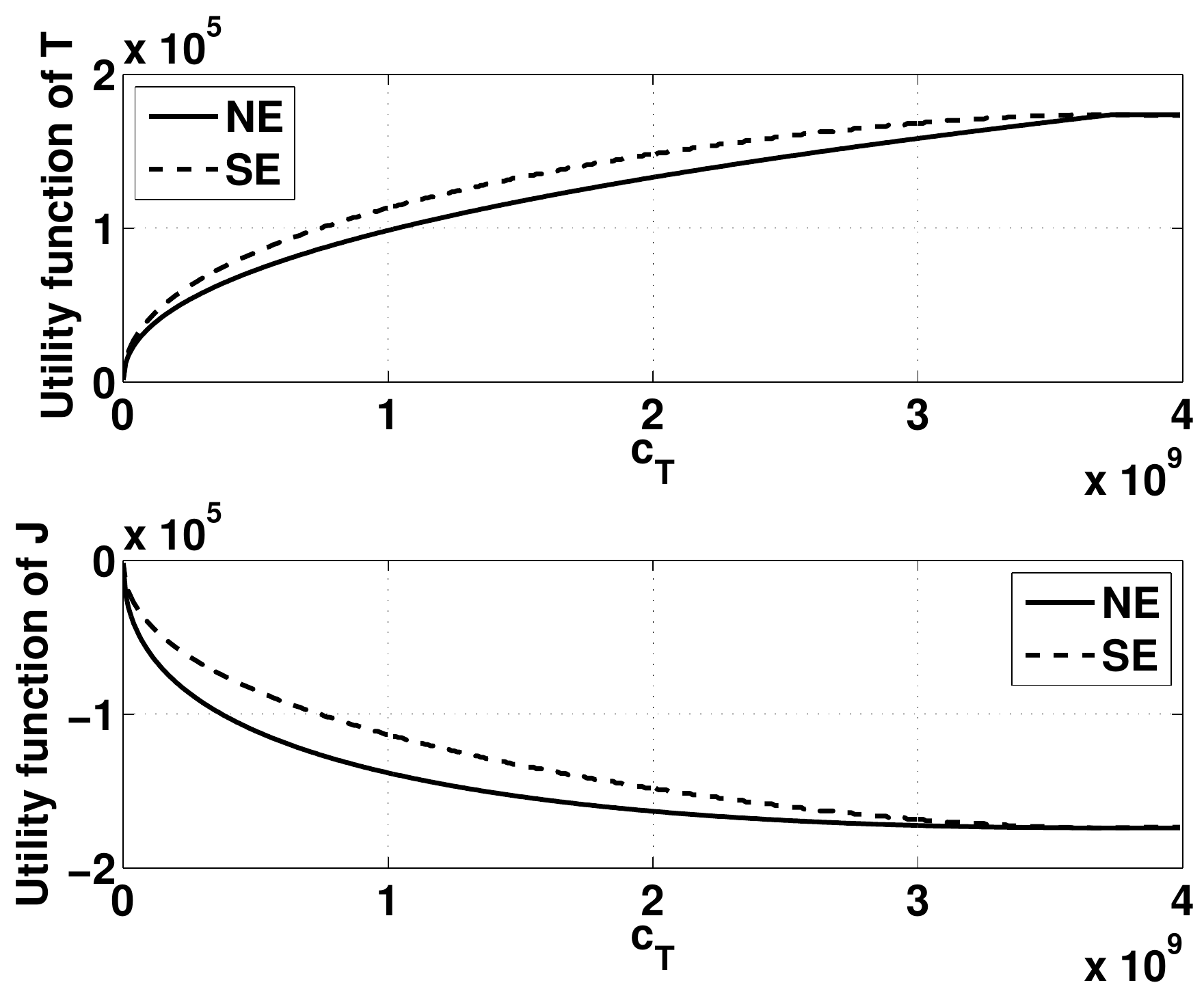}
   \caption{Comparison between the utilities achieved by each player at the NE and SE as a function of the weight parameter $c_T$ ($c_{T^*} \cdot P = 2 \cdot 10^6$).}
   \label{payoffsSE}
\end{figure}

We now turn to the analysis of the Stackelberg game, where the target node anticipates the jammer's reaction. In this regard,
Fig. \ref{payoffsSE} compares the utilities achieved by each player at the NE and SE. 
Note that, as proven in Theorem \ref{theorem:improvement}, the utility achieved by the
target node at the SE is higher than, or at least equal to, the utility achieved at the NE. 
Moreover, at the SE the utility is higher than at the NE for the jammer as well. 
In fact, the target node increases the maximum silence duration $x$, that is, it increases transmission delay, and inhibits the jammer. 
Accordingly, the jammer stops its disrupting attack, and thus, it saves energy;
as a result, its utility increases when compared to that at the Nash Equilibrium. 
We further observe that, as expected, for high values of $c_T$, the improvement in the achieved utility becomes negligible, as already proven in 
Theorem \ref{theorem:improvement}. 

\begin{figure}
\centerline{
\subfigure[]{
\includegraphics[width=.45\textwidth]{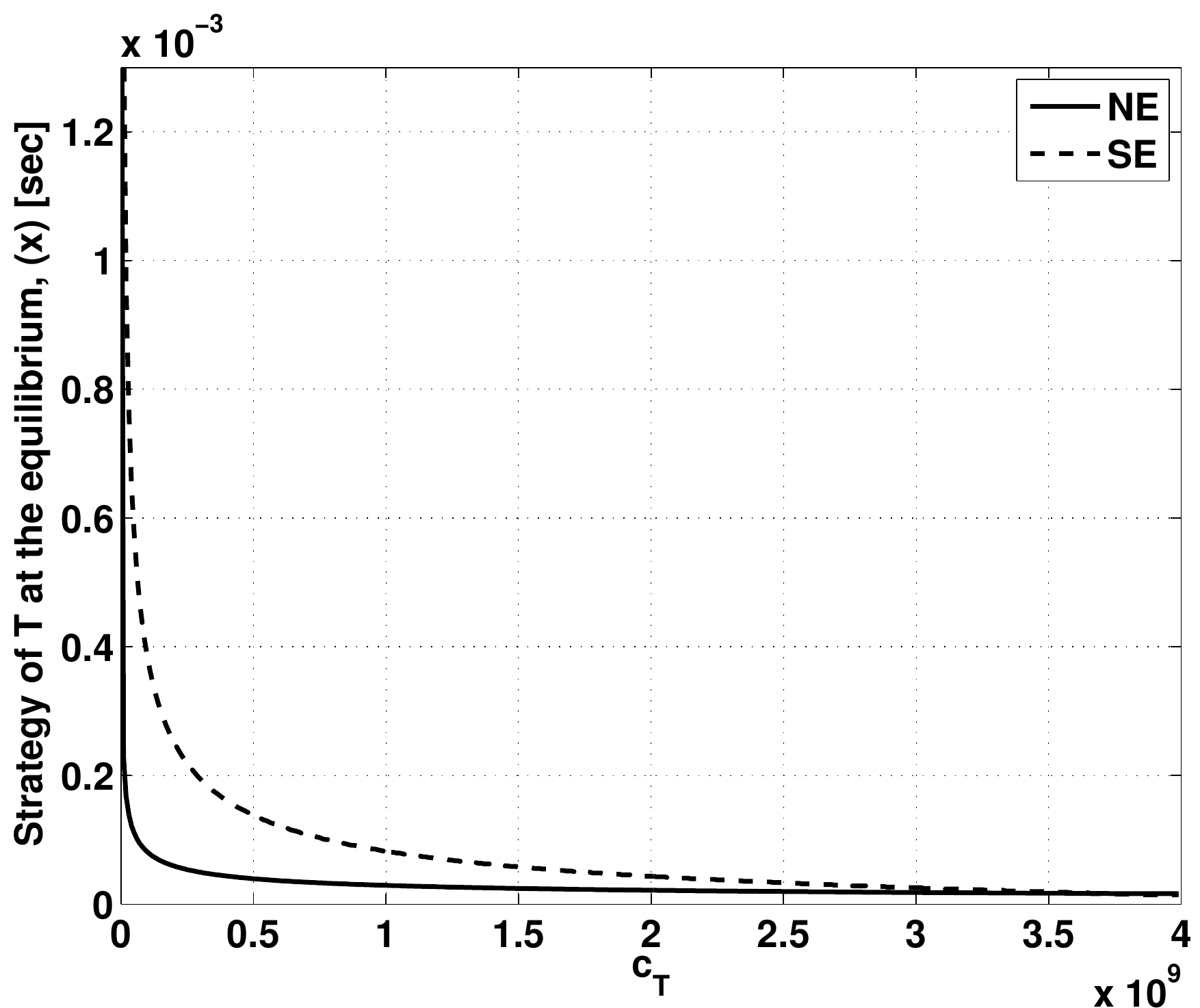}\label{SEx}
}
\hfil
\subfigure[]{
\includegraphics[width=.45\textwidth]{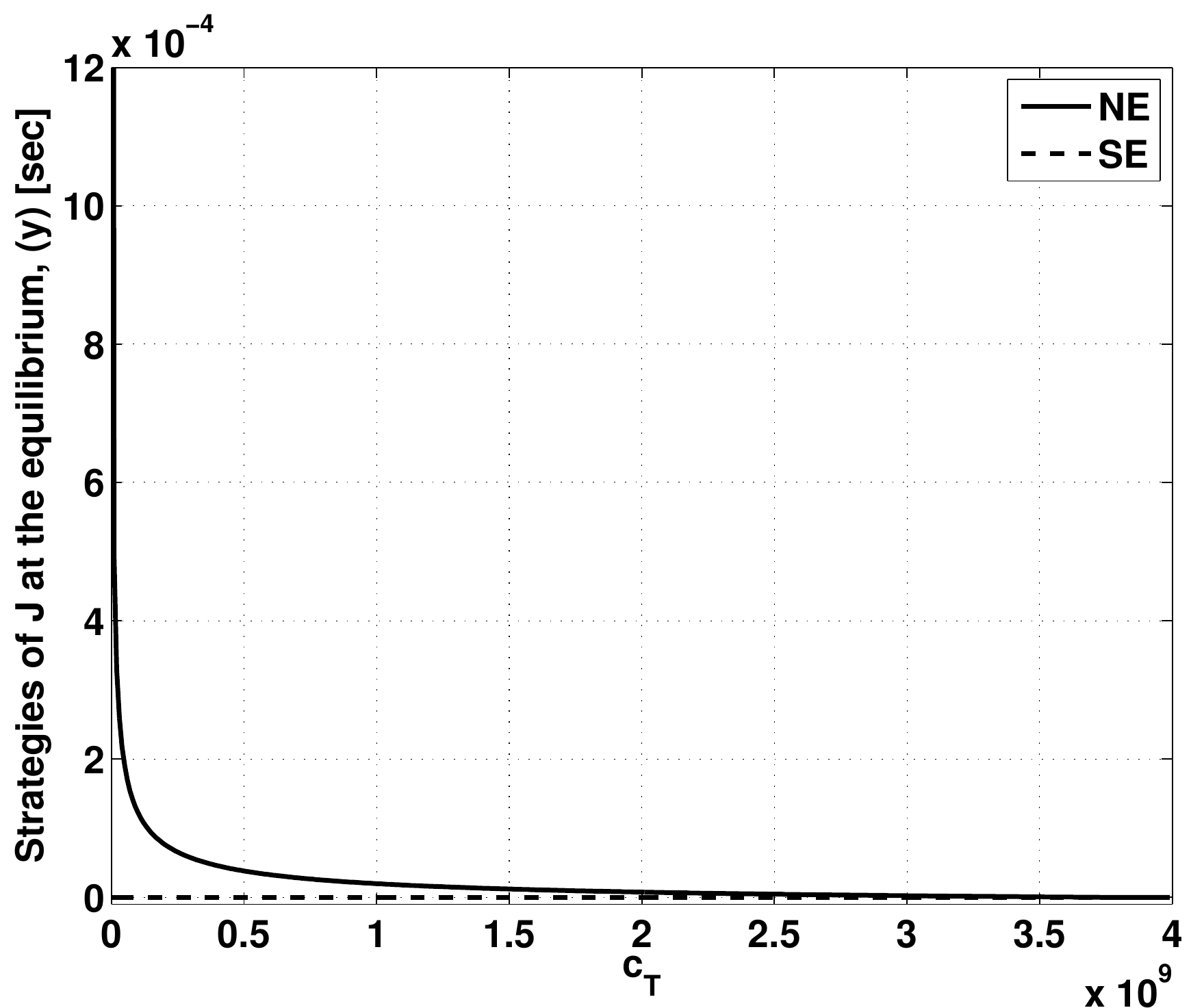}\label{SEy}
}
}
\caption{a) Strategy of the target node $T$ at NE and SE as a function of the weight parameter $c_T$ b) Strategy of the jammer $J$ at NE and SE as a function of the weight parameter $c_T$.}
\label{}
\end{figure}

Figs. \ref{SEx} and \ref{SEy} illustrate the strategy at the equilibrium points of the target node and the jammer
as a function of the parameter $c_T$, and show how the strategies of both players decrease as $c_T$ increases.
In fact, high values of the weight parameter $c_T$ suggest a conservative behavior of the jammer at the NE (e.g. the jammer is more energy constrained),
so that the jammer prefers to decrease the duration of the jamming signal $y$ in order to reduce its energy consumption.
Instead, as proven in Theorem \ref{theorem:SET},  at the SE the target node forces the jammer in stopping its jamming attack, thus,  $y_{\mathrm{SE}}=0$.
Furthermore, for high values of the parameter $c_T$, the strategy $x$ of the target node consists in choosing low silence duration at both the NE and SE.
This is because by increasing $c_T$
the strategy of the jammer consists in reducing the duration of the jamming signal. Hence, the target node decreases the maximum duration of the silence intervals $x$, 
that is, T reduces the transmission delays 
while achieving a higher transmission capacity. Note that when the value of $c_T$ approaches $\tilde{c_T}$, the NE and SE become equal.

\begin{figure}[t]
 \centering
 \subfigure[]{\includegraphics[width=0.4\textwidth]{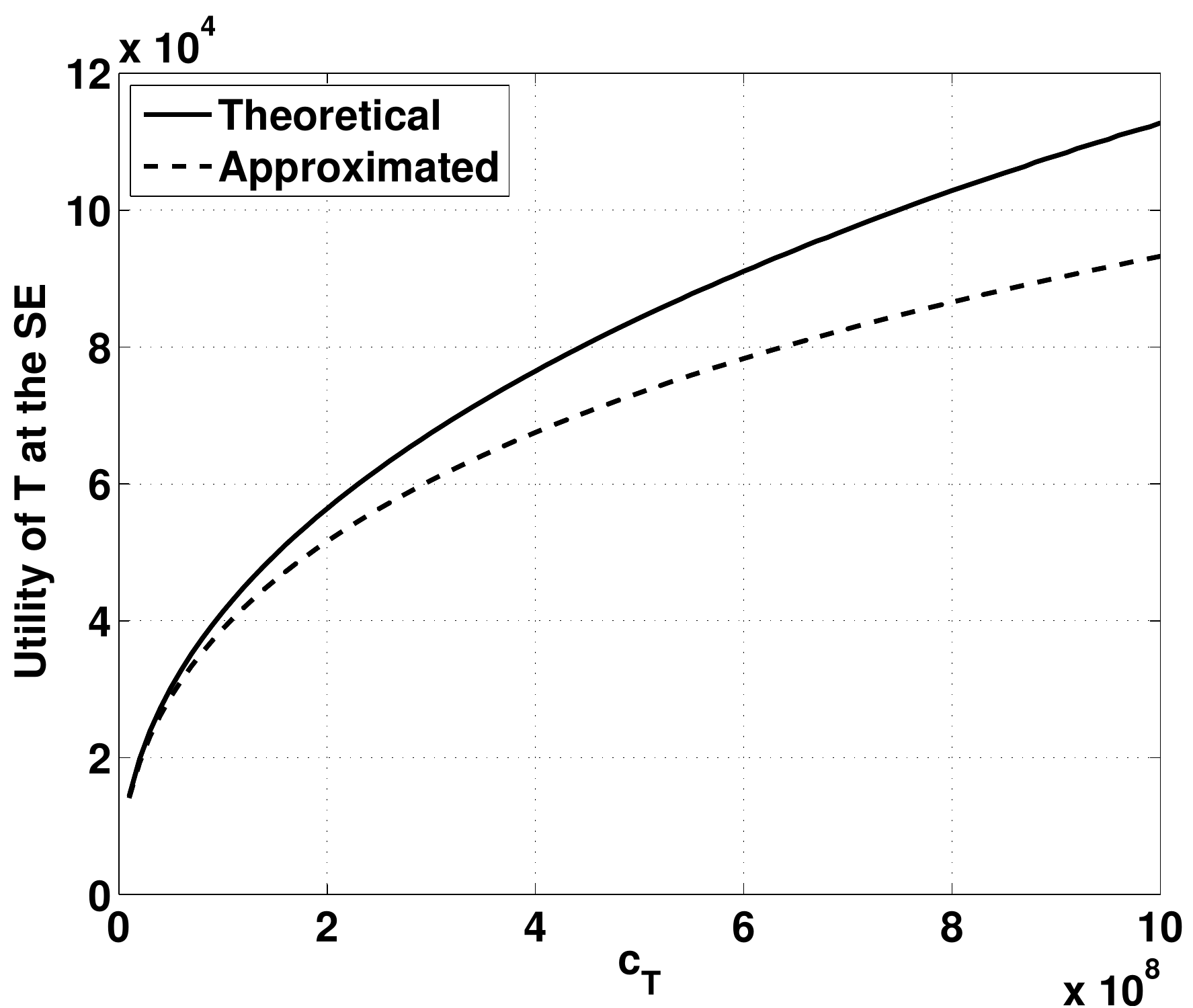}
 \label{approximationUtility}
 }
 \centering
 \subfigure[]{\includegraphics[width=0.4\textwidth]{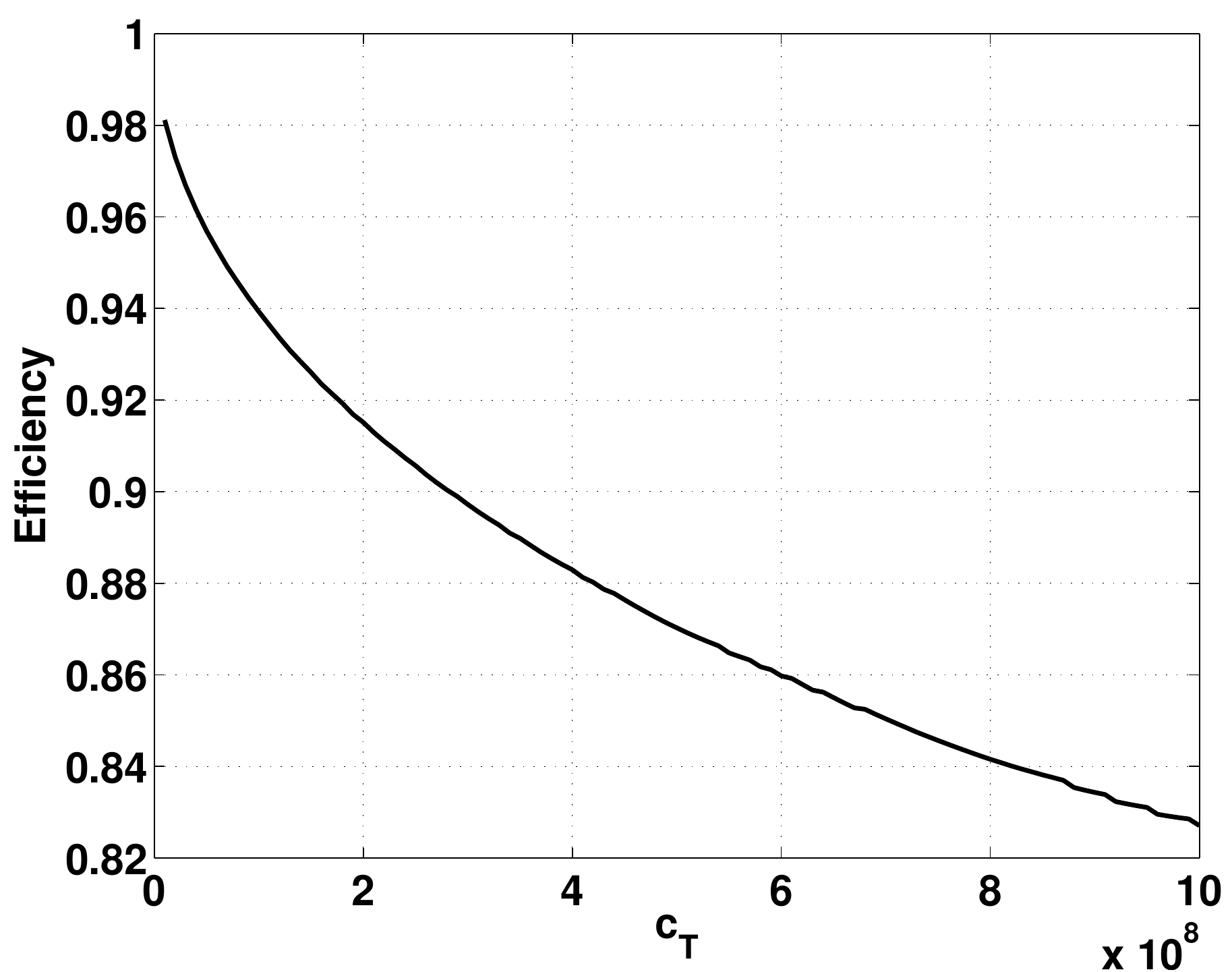}
 \label{approximationEfficiency}
 }
 \centering
 \subfigure[]{\includegraphics[width=0.4\textwidth]{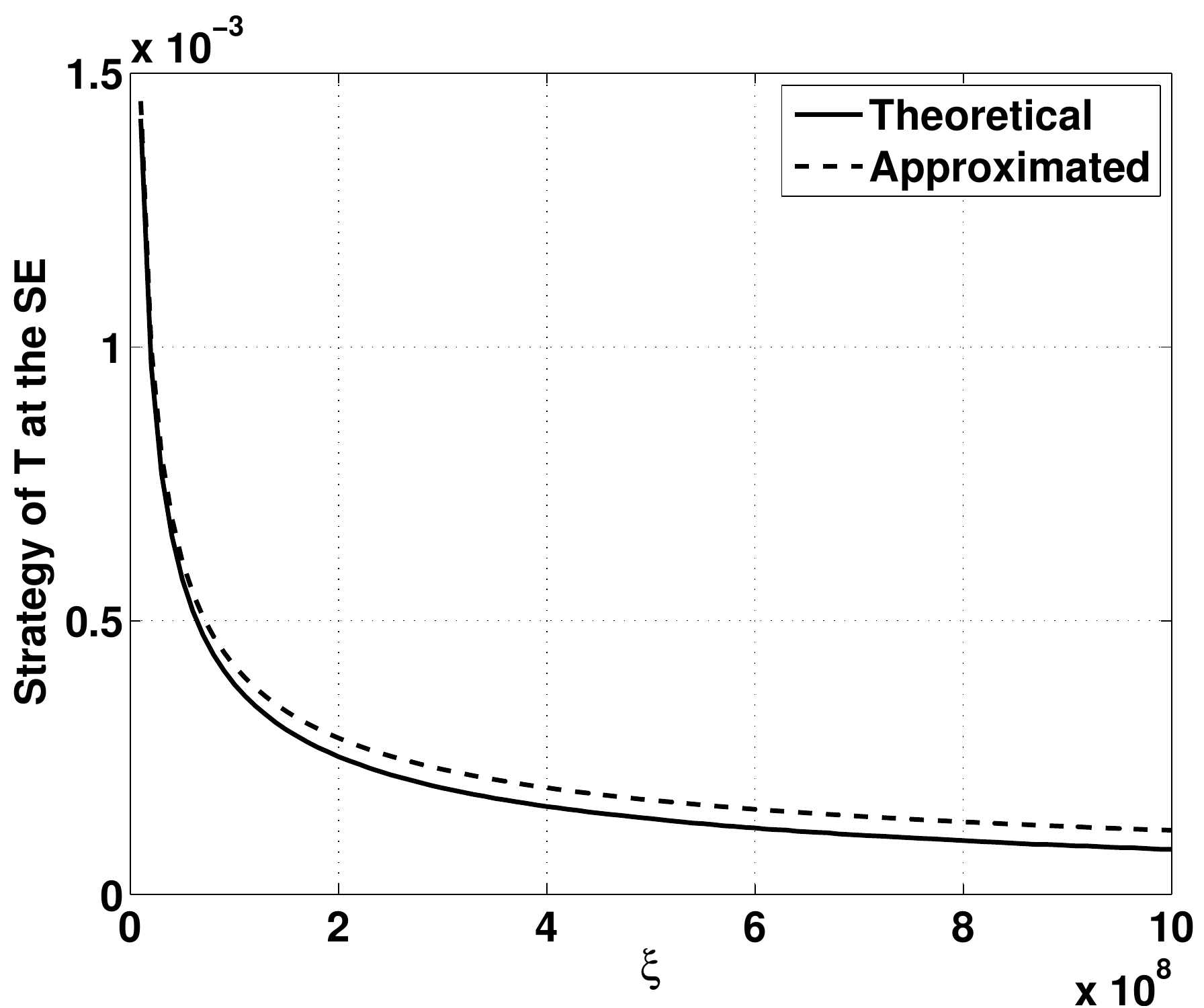}
 \label{approximationX}
 }
 \caption{Impact of the approximation $x_{\mathrm{SE}}'$ in eq. (\ref{eq:xTildeApproxFinal}) on the Stackelberg game outcome as a function of the weight parameter $c_T$ ($c_{T} \cdot P = 2 \cdot 10^6$).}
\end{figure}

Under the perfect knowledge assumption, at the SE the strategy of the target node, $x_{\mathrm{SE}}$, coincides with the solution of $\chi(x)=0$, which can also be approximated 
to $x_{\mathrm{SE}}'$ as given in eq. (\ref{eq:xTildeApproxFinal}).
Accordingly, in Fig. \ref{approximationUtility} we compare the utilities of the target node at the SE, in its exact and 
approximated strategies $x_{\mathrm{SE}}$ and $x_{\mathrm{SE}}'$, respectively.
Fig. \ref{approximationEfficiency} shows that the approximation accuracy of $x_{\mathrm{SE}}'$, defined as the ratio between $\mathcal{U}_{T}(x_{\mathrm{SE}}',b_{J}(x_{\mathrm{SE}}'))$ and
$\mathcal{U}_{T}(x_{\mathrm{SE}},b_{J}(x_{\mathrm{SE}}))$, strongly depends on the value of $c_T$. 
As shown in Fig. \ref{approximationX}, the error introduced by the approximation 
$\left( T_{AJ}+\frac{x_{\mathrm{SE}}}{2}\right) \approx \frac{x_{\mathrm{SE}}'}{2}$ is low
when low values of $c_T$ are considered,
because, in this case, the strategy of T at 
the SE, $x_{\mathrm{SE}}$, consists in choosing larger silence durations, and thus $\frac{x_{\mathrm{SE}}}{2} \gg T_{AJ}$.
\textcolor{black}{On the contrary}, when $c_T$ is high, there is no need for the target node to choose high $x_{\mathrm{SE}}$ values, thus the 
above approximation introduces a non-negligible error on the estimate of
$x_{\mathrm{SE}}'$. Note that, although the approximation is affected by errors, Fig. \ref{approximationEfficiency} shows that the approximation accuracy
is still high (\textcolor{black}{i.e. larger than 82\%}).

\begin{figure}[t]
\centering
   \includegraphics[width=0.5\columnwidth]{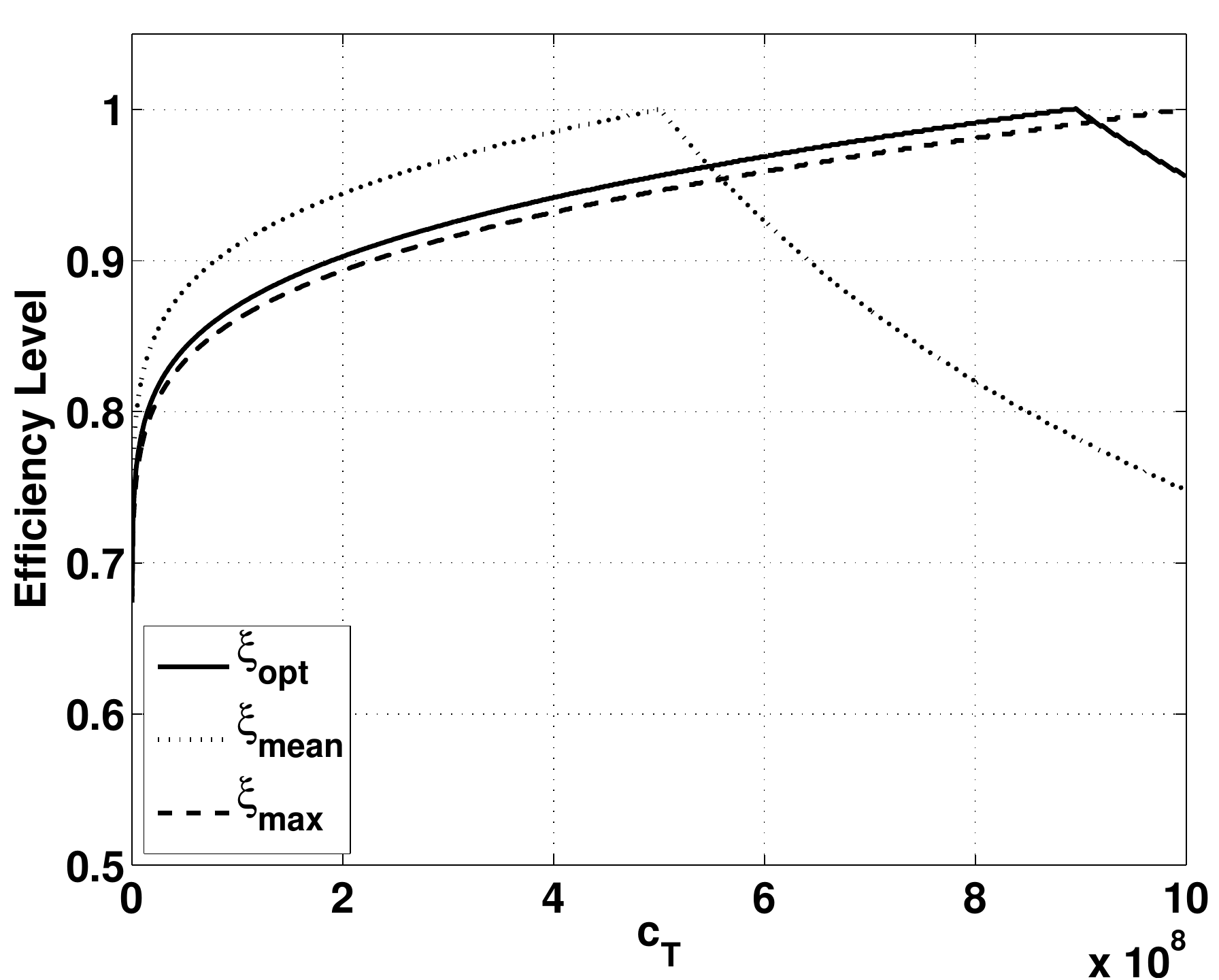}
   \caption{Equilibrium efficiency $e(\xi)$ as a function of the weight parameter $c_T$ ($c_{T} \cdot P = 2 \cdot 10^6$).}
   \label{efficiency}
\end{figure}

To evaluate the impact of imperfect knowledge on the utility of the target node, let us now
define the \emph{equilibrium efficiency} $e(\xi)$ as follows:
\begin{equation}
 e(\xi)=\frac{{\bf U}_{T}^{\xi}}{{\bf U}_{T}^{c_T}}
\end{equation}
Fig. \ref{efficiency} illustrates the equilibrium efficiency of the target node as a function of $c_T$ for different choices of $\xi$.
More in detail, we considered $\xi \in \{ \xi_{opt},\xi_{mean},\xi_{max},\xi_{min}$ \}, where $\xi_{mean}=(\xi_{max}+\xi_{min})/2$, $\xi_{min}=10^5$ and $\xi_{max}=10^9$.
Note that in our simulations $\xi_{min}=10^5$ and $\xi_{max}=10^9$ are realistic setting assumptions. 
In fact, lower values of $\xi_{min}$ or higher values of $\xi_{max}$ lead to unbalanced settings as one of the terms in eq. (\ref{utility_scm}) will always dominate the other.
The most important result is that the equilibrium efficiency when $\xi \in \{ \xi_{opt},\xi_{mean},\xi_{max} \}$ is always higher than 75\%, while the case $\xi=\xi_{min}$ 
achieves a very low equilibrium efficiency (and thus, it is not reported in Fig. \ref{efficiency}).
As demonstrated in Section \ref{sec:imperfect}, Fig. \ref{efficiency} shows that $\xi_{max}$ well approximates $\xi_{opt}$, i.e., $e(\xi_{opt})\simeq e(\xi_{max})$.
Therefore, from a practical point of view, if the
computation of $\xi_{opt}$ is not feasible (e.g., high computational cost and 
low hardware capabilities) it is still possible to achieve a high equilibrium efficiency by choosing $\xi=\xi_{max}$.

\begin{figure}[htb!]
\centering
  \includegraphics[scale=0.55]{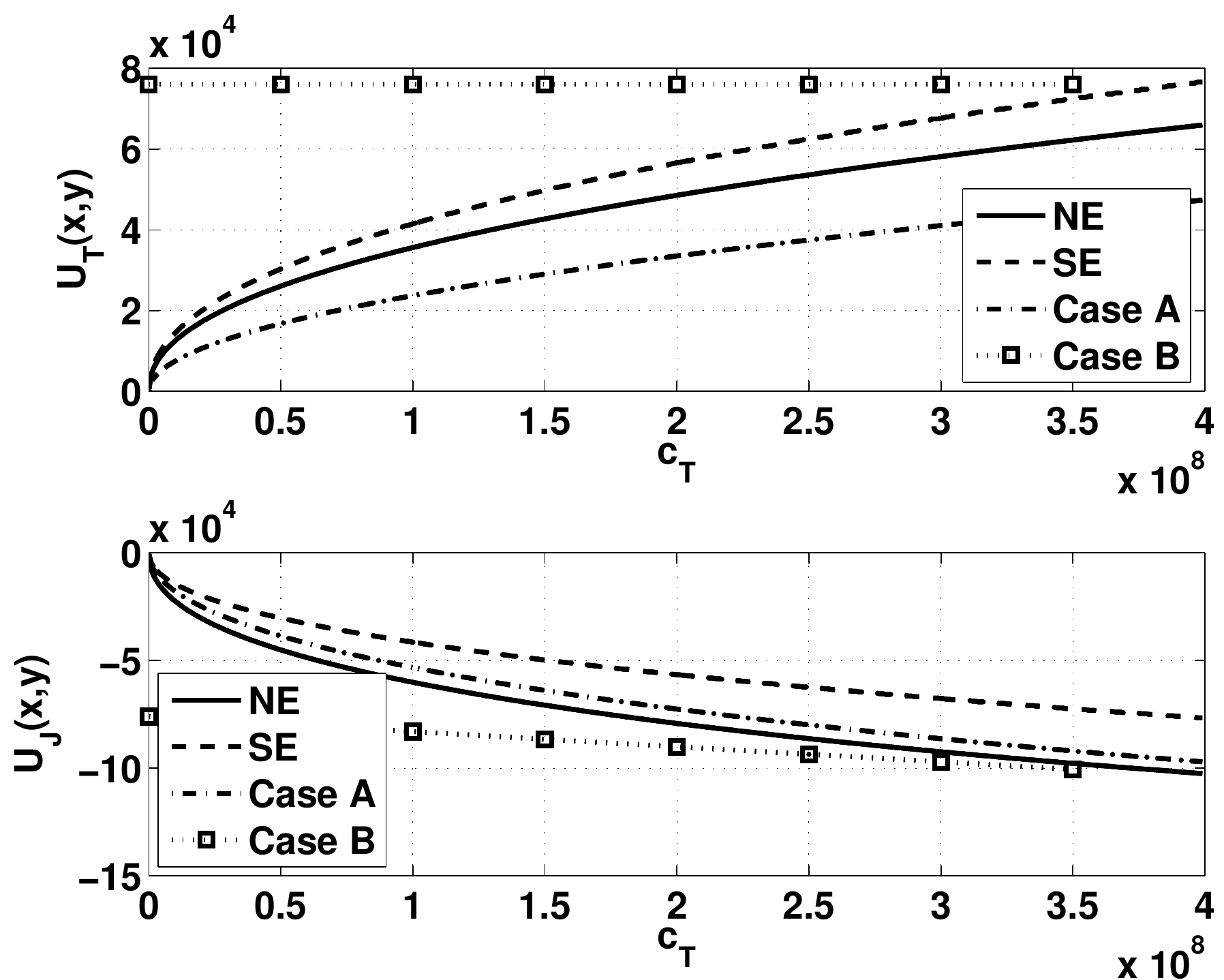}
  \caption{Comparison between the utility of the target node and the jammer when they work at the NE, at the SE and what is obtained in Case A and B.}
  \label{comparison}
\end{figure}

\textcolor{black}{
Finally, in Fig. \ref{comparison} we compare the utility functions of the target node and the jammer obtained at the NE and SE with what is obtained in the cases the two players select their strategies without considering the strategies of each other.
More specifically we will consider the two following cases:
\begin{itemize}
	\item
		{\it Case A}: The target node selects its strategy $x$ in such a way that its capacity is maximized without considering that the jammer will try to disrupt the communication in the timing channel as well.
In other terms, the target node will assume that $y \approx 0$.
	\item
		{\it Case B}: The jammer selects its strategy $y$ assuming that the target node is not aware that it (the jammer itself) is trying to disrupt the communication in the timing channel.
In other terms, the jammer will assume that $x \approx b_T(0)$.
\end{itemize}
}

\textcolor{black}{
When compared to the NE and SE cases the utility function of the target node will decrease in Case A and increase in Case B.
The viceversa holds for the utility function of the jammer.
We observe that the gap between the utility functions obtained in Cases A and B compared to the NE and SE decrease when the cost $c_T$ increases.
This is because when the cost $c_T$ increases the jammer becomes more concerned about the energy consumption and therefore the value $y_{\mathrm{NE}}$ becomes smaller.
Accordingly, the assumptions considered in Cases A and B become accurate and consequently the behavior approaches what is obtained when each player takes the strategy of the opponent into account.
}

\subsection{Simulation results}
\textcolor{black}{To assess the accuracy of the theoretical results derived in the previous sections, we implemented a simulator that shows how players' behavior dynamically evolves and how players choose their strategies. 
In the  simulations we assume that each player chooses its own initial strategy randomly. 
\begin{figure}[t]
\centering
   \includegraphics[width=0.5\columnwidth]{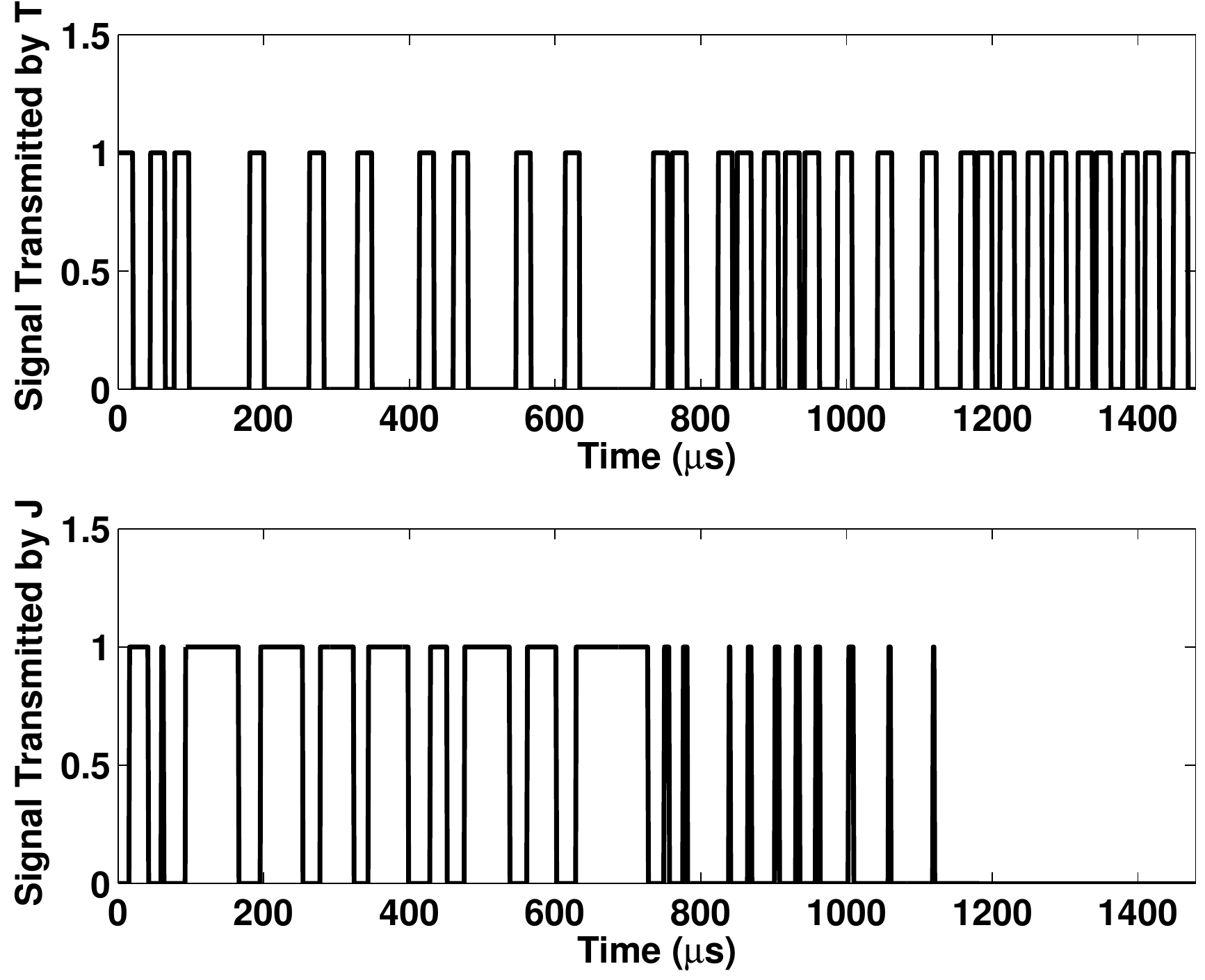}
   \caption{Strategies chosen by the players vs. time.}
   \label{sim_strategy}
\end{figure}
Then, players update their strategies each 10 cycles
 during which each player estimates the opponent's strategy.
Players update their strategies according to the BRD discussed in Section \ref{sec:convergence}.  
The simulation parameter setup is summarized in Table \ref{SimulationSetup}. Note that we chose $c_T>c_T^{max}$ so that NE is on the border, i.e., the strategy of the jammer at the NE is $y^*=0$.
\begin{table}[t]
  \begin{center} 
    \begin{tabular}{|l|l|l|}
	    \hline
	    Name & Value & Unit\\
	    \hline
            $T_{AJ}$ & 15 & $\mu$s \\
	    \hline
	    $\Delta$ & 1 & $\mu$s \\
	    \hline
	    $P$ & 2 & W \\
	    \hline
        $T_{P}$ & 20 & $\mu$s \\
	    \hline
        $c_{T}$ & $8 \cdot 10^9$ & $bit/(sec \cdot J)$ \\
	    \hline
        $c_{T^*}$ & $10^6$ & $bit/(sec \cdot J)$ \\
	    \hline
    \end{tabular}
    \end{center}
    \caption{\label{SimulationSetup} Parameter settings used in our simulations.}
\end{table}
In Fig. \ref{sim_strategy} we show an example of the simulation results that illustrates how players dynamically change their strategies depending on the opponent's one. The figure shows that after three iterations, players reach the NE, that is, due to the high energy cost, the jammer stops its attack while the target node chooses its strategy according to its best response function,i.e., $x^*=b_T(0)$. 
}

%% file: ConcBib.tex
\section{Conclusions}  \label{conclusions}

In this paper we have proposed a game-theoretic model of the interactions between a jammer and a communication node that exploits
a timing channel to improve resilience to
jamming attacks.
Structural properties of the utility functions of the two players have been analyzed and exploited to prove the existence and uniqueness
of the Nash Equilibrium.
The convergence of the game to the Nash Equilibrium has been studied and proved by analyzing the best response dynamics.
Furthermore, as the reactive jammer is assumed to start transmitting its interference signal only after detecting activity of the node under attack, a
Stackelberg game has been properly investigated, and proofs on the existence
and uniqueness of the Stackelberg Equilibrium has been provided.
Finally, the case of imperfect knowledge about the parameter $c_T$ has been also discussed.
Numerical results, derived in several real network settings, show that our proposed models well capture the main factors behind the utilisation of timing channels, 
thus representing a promising framework for the design and understanding of such systems.

%% file: bib.bbl

%% file: appendixit.tex
\section{Proof of Theorem \ref{theorem:NEunique}} \label{app:NEunique}

    \begin{IEEEproof}
            In order to prove the theorem we have to solve eq. (\ref{bestrespInter}), that is, find a pair $(x,y)$ which solves the following system of equations:
            \begin{equation}
             \begin{cases}
              y=\chi(\Delta e^{ \psi(y) +1} ) \\ 
              x=\Delta e^{ \psi(y) +1} 
             \end{cases}
             \label{SystemEqInt}
            \end{equation}
            \noindent
            By exploiting the Lambert W-function definition and the relationship $z/W(z)=e^{W(z)}$, where $z=\left[ \frac{2(T_{AJ}+y)}{e\Delta}\right]$,
            it can be proven that the above system leads to
            \begin{equation}
              (y+T_{AJ})^2=\frac{1}{\eta} \cdot \frac{\psi^2(y)}{\psi(y) +1} 
              \label{beforeDerivative}
            \end{equation}
            \noindent
            Given that the first derivative of the Lambert W-function is defined as
            \begin{equation}
              W'(z)=\frac{W(z)}{z(W(z)+1)}
              \label{Wfirstderivative}
            \end{equation}
            \noindent
            eq. (\ref{beforeDerivative}) can also be rewritten as
            \begin{equation}
             e^{W\left( \frac{2(T_{AJ}+y)}{e \Delta} \right)} = \frac{1}{\eta} \cdot \frac{2}{\Delta e} \cdot W'\left( \frac{2(T_{AJ}+y)}{e \Delta} \right)
             \label{EqFinal}
            \end{equation}
            \noindent
            Note that the function on the left-hand side is strictly increasing, while the one on the right-hand side is strictly decreasing.
            These structural properties imply that 
            the two functions have no more than one intersection point.
            Therefore, the game admits a unique NE.

            Now we focus on finding a closed form for the unique NE.

            To this purpose, eq. (\ref{EqFinal}) can be reformulated as
             \begin{equation}
              e^{2 W\left( \frac{2(T_{AJ}+y)}{e \Delta} \right)} (W\left( \frac{2(T_{AJ}+y)}{e \Delta} \right)+1) =  \frac{1}{\eta}\left(\frac{2}{e \Delta} \right)^2
              \label{equseful}
              \nonumber
             \end{equation}
             \noindent
             which, by exploiting the relation $z=W(z)e^{W(z)}$, can be rewritten as follows:
             \begin{equation}
              W\left( \frac{2(T_{AJ}+y)}{e \Delta} \right) = \frac{1}{2}W\left( \frac{8}{\eta \Delta^2} \right)-1
              \label{y_ne_inter}
             \end{equation}
             \noindent
             It is easy to prove that eq. (\ref{y_ne_inter}) has the following solution 
             \begin{equation}
              y^*=\frac{\Delta}{2}\left(\frac{1}{2}W\left( \frac{8}{\eta \Delta^2} \right)-1\right)e^{\frac{1}{2}W\left( \frac{8}{\eta \Delta^2} \right)}-T_{AJ}
              \label{y_ne_final}
             \end{equation}
             \noindent
             By substituting eq. (\ref{y_ne_final}) in eq. (\ref{bestTC}) we obtain $x^*=\Delta e^{\frac{1}{2}W(\frac{8}{\eta \Delta^2})}$.
             As the point $(x^*,y^*)$ has been obtained as the intersection between the best response functions in eqs. (\ref{bestTC}) and (\ref{bestJ}), 
             it follows that $(x_{\mathrm{NE}},y_{\mathrm{NE}})=(x^*,y^*)$ is the unique NE.

             Finally, we prove that the NE $(x_{\mathrm{NE}},y_{\mathrm{NE}})$ is an \emph{interior} NE.
             An interior NE happens when it is not on the border of the strategy set; therefore, we aim at proving that $x_{\mathrm{NE}} >2\Delta$ and $y_{\mathrm{NE}} > 0$.
             As $x_{\mathrm{NE}}=\Delta e^{\frac{1}{2}W(\frac{8}{\eta \Delta^2})}$, proving that $x_{\mathrm{NE}}$ is not on the border is trivial;  from eq. (\ref{y_ne_final})
             it can also be easily proven that the condition
             $y_{\mathrm{NE}} > 0$ implies $0<c_T < \tilde{c_T}$, \textcolor{black}{where $\tilde{c_T}$ is given in eq. (\ref{cTTilde});}
              therefore, an interior NE exists only if $0< c_T < \tilde{c_T}$.
             Theorem \ref{theorem:NEexist} states that an NE must exist for any given weight parameter $c_T$. 
             Since we already proved that an interior NE exists only if $0< c_T < \tilde{c_T}$, we can deduce that the NE is on the border if $c_T \geqslant \tilde{c_T}$ .
             
             From eq. (\ref{bestJ}) we know that for $c_T \geqslant \tilde{c_T}$ the best response function of the jammer, 
             $b_{J}(x)$, is continuous, and it is upper-bounded by $b_{J}(\hat{x})$ where $\hat{x} = \Delta e^{\frac{1}{2}W(\frac{2}{\eta \Delta^2})}$, 
             and lower-bounded by $0$; thus, as the NE has to be at the border, 
             it follows that the only feasible solution is $y_{\mathrm{NE}} = 0$.
             Hence, from eqs. (\ref{bestTC}) and (\ref{bestJ}), it is easy to derive closed form solutions on the border NE,
             $(x_{\mathrm{NE}},y_{\mathrm{NE}})= \left(\Delta e^{W(\frac{2T_{AJ}}{e\Delta})+1} , 0\right)$, which concludes the proof.
    \end{IEEEproof}

\section{Proof of Lemma \ref{LemmaConditionOnBothXandCt}} \label{app:lemmaConvergence}

	\begin{IEEEproof}
To prove the Lemma, it will be shown that the condition in eq. (\ref{Lemma2Conclusion}) implies that the Jacobian matrix norm $|| J_\textbf{b} ||_{\infty}$ in eq. (\ref{jacobian}) is lower than 1.
In fact, the condition $|| J_\textbf{b} ||_{\infty} < 1$ leads to: 
\begin{equation}
\max \left(\left |\frac{\partial}{\partial y} b_T(y)\right | , \left |\frac{\partial}{\partial x} b_J(x)  \right | \right)< 1
\nonumber
\end{equation}
 
 Note that $\left| \frac{\partial}{\partial y} b_T(y) \right|$ can be calculated as 
 \begin{align}
  \begin{array}{rcl}
   \left| \frac{\partial}{\partial y} b_T(y) \right|  & = & \frac{2}{W(\frac{2(T_{AJ}+y)}{e \Delta})+1} \nonumber \\ 
  \end{array}
 \end{align}
 \noindent

The above function is non-negative and strictly decreasing, thus it achieves its maximum value when $y=0$.
Accordingly, it is sufficient to show that  
 \begin{equation}
  \max_{y \in \mathcal{S_J}} \left(\frac{2}{W(\frac{2(T_{AJ}+y)}{e \Delta})+1}\right)  < 1  \hspace{0.5cm},\hspace{0.5cm} \forall y\geq0
  \nonumber
 \end{equation}
 \noindent
or, equivalently, that
 \begin{equation}
  \max_{y \in \mathcal{S_J}} \left(\frac{2}{W(\frac{2(T_{AJ}+y)}{e \Delta})+1}\right) =\frac{2}{W(\frac{2T_{AJ}}{e \Delta})+1}  < 1  \hspace{0.5cm},\hspace{0.5cm} \forall y\geq0
  \label{eqBTder}
  \nonumber
 \end{equation}
 \noindent
which is indeed satisfied for all values of $y$ in the strategy set;
therefore, $\left| \frac{\partial}{\partial y} b_T (y)\right| < 1,  \forall y \in \mathcal{S_J}$.

\textcolor{black}{Concerning the condition $\left| \frac{\partial}{\partial x} b_J(x) \right| < 1$, by deriving $b_J(x)$, it follows that  }
\begin{equation} \label{equazione}
\left| \frac{1}{2} \left [  \frac{1}{x \sqrt{\eta \log \frac{x}{\Delta}}}-1 \right] \right| < 1 
\end{equation}

\textcolor{black}{The expression on the right-hand side of eq. (\ref{equazione}) is a non-negative strictly decreasing function, so again
eq. (\ref{equazione}) results in
}

\begin{equation}
 \max_{x \in \mathcal{S_T}} \left( \left| \frac{1}{2} \left[\frac{1}{x\sqrt{ \eta \log\left( \frac{x}{\Delta} \right)}}-1 \right] \right| \right) < 1 \label{eqmax}
\end{equation}
Note that eq. (\ref{eqmax}) can be rewritten in the form given in eq. (\ref{Lemma2Conclusion})
\textcolor{black}{and $|| J_\textbf{b} ||_{\infty} =|| J_\textbf{b} ||$ as $J_\textbf{b}$ is diagonal. Let $s^i=(x^i,y^i)$, it then follows that 
\begin{eqnarray*}
||s^{i+1}-s^i||\le ||J_\textbf{b}^{max}|| \cdot ||s^i-s^{i-1}||\le \cdots \le ||J_\textbf{b}^{max}||^i ||s^1-s^0||
\end{eqnarray*}
where $||J_\textbf{b}^{max}|| =\max J_\textbf{b}$.
The above equation indicates that given any $\epsilon>0$, after at most $\log_{J_\textbf{b}^{max}}\frac{\epsilon}{||s^1-s^0||}$ iterations, the game converges to the NE as $||s^{i+1}-s^i||\le \epsilon$ which thus concludes the proof.}
\end{IEEEproof}